\documentclass[%
 reprint,
 amsmath,amssymb,
 aps,
]{revtex4-2}

\usepackage{graphicx}
\usepackage{dcolumn}
\usepackage{bm}

\usepackage[utf8]{inputenc}
\usepackage[T1]{fontenc}
\usepackage{mathptmx}
\usepackage{etoolbox}
\usepackage{float}
\usepackage{hyperref} 
\hypersetup{colorlinks,allcolors=black}
\usepackage{xfrac}
\usepackage{textcomp}
\usepackage{mathpazo}
\usepackage{verbatim}
\usepackage{color,soul}

\definecolor{gray}{rgb}{0.8, 0.83, 0.9}

\usepackage{scalerel}
\usepackage{xcolor}
\usepackage{tikz}
\usepackage{times}
\usetikzlibrary{svg.path}
\definecolor{orcidlogocol}{HTML}{A6CE39}
\tikzset{
  orcidlogo/.pic={
    \fill[orcidlogocol] svg{M256,128c0,70.7-57.3,128-128,128C57.3,256,0,198.7,0,128C0,57.3,57.3,0,128,0C198.7,0,256,57.3,256,128z};
    \fill[white] svg{M86.3,186.2H70.9V79.1h15.4v48.4V186.2z}
                 svg{M108.9,79.1h41.6c39.6,0,57,28.3,57,53.6c0,27.5-21.5,53.6-56.8,53.6h-41.8V79.1z M124.3,172.4h24.5c34.9,0,42.9-26.5,42.9-39.7c0-21.5-13.7-39.7-43.7-39.7h-23.7V172.4z}
                 svg{M88.7,56.8c0,5.5-4.5,10.1-10.1,10.1c-5.6,0-10.1-4.6-10.1-10.1c0-5.6,4.5-10.1,10.1-10.1C84.2,46.7,88.7,51.3,88.7,56.8z};
  }
}

\newcommand\orcid[1]{\href{https://orcid.org/#1}{\mbox{\scalerel*{
\begin{tikzpicture}[yscale=-1,transform shape]
\pic{orcidlogo};
\end{tikzpicture}
}{|}}}}

\makeatletter
\def\@email#1#2{%
 \endgroup
 \patchcmd{\titleblock@produce}
  {\frontmatter@RRAPformat}
  {\frontmatter@RRAPformat{\produce@RRAP{*#1\href{mailto:#2}{#2}}}\frontmatter@RRAPformat}
  {}{}
}%

\makeatother

\begin{document}

\preprint{APS/123-QED}

\title{
Electrical Tuning of Terahertz Plasmonic Crystal Phases
}


\author{
\orcid{0000-0002-6143-9469}
P.~Sai$^{1},$
\orcid{0000-0001-5629-8545}
V.~V.~Korotyeyev$^{2},$
\orcid{0000-0002-6898-4638}
M.~Dub$^{1},$
\orcid{0000-0001-7265-3199}
M.~S\l{}owikowski$^{1,3},$
M.~Filipiak$^{1,3},$
\orcid{0000-0002-0735-4608}
D.~B.~But$^{1},$
\orcid{0000-0002-7444-4056}
Yu.~Ivonyak$^{1},$
\orcid{0000-0002-7407-6492}
M.~Sakowicz$^{1},$
\orcid{0000-0001-9112-3505}
Yu.~M.~Lyaschuk$^{2},$
\orcid{0000-0001-6901-2657}
S.~M.~Kukhtaruk$^{2},$
\orcid{0000-0003-3616-8756}
G.~Cywi\'nski$^{1},$
\orcid{0000-0003-4537-8712}
W.~Knap$^{1}$}

\email{psai@unipress.waw.pl}

             
\affiliation{\vspace{0.3cm}$^1$CENTERA Laboratories, Institute of High Pressure Physics PAS, Warsaw, Poland}

\affiliation{$^2$V.~Ye.~Lashkaryov Institute of Semiconductor Physics (ISP), NASU, Kyiv, 03028, Ukraine}

\affiliation{$^3$Centre for Advanced Materials and Technologies, Warsaw University of Technology, Warsaw, Poland}


\begin{abstract}
We present an extensive study of resonant two-dimensional (2D) plasmon excitations in grating-gated quantum well heterostructures, which enable an electrical control of periodic charge carrier density profile.
Our study combines theoretical and experimental investigations of nanometer-scale AlGaN/GaN grating-gate structures and reveals that all terahertz (THz) plasmonic resonances in these structures can be explained only within the framework of the plasmonic crystal model.
We identify two different plasmonic crystal phases. 
The first is the \textit{delocalized} phase, where THz radiation interacts with the entire grating-gate structure that is realized at a weakly modulated 2D electron gas (2DEG) regime. 
In the second, the \textit{localized} phase, THz radiation interacts only with the ungated portions of the structure.
This phase is achieved by fully depleting the gated regions, resulting in strong modulation.
By gate-controlling of the modulation degree, we observe a continuous transition between these phases.
We also discovered that unexpectedly the resonant plasma frequencies of ungated parts (in the \textit{localized} phase) still depend on the gate voltage.
We attribute this phenomenon to the specific depletion of the conductive profile in the ungated region of the 2DEG, the so-called edge gating effect.
Although we study a specific case of plasmons in AlGaN/GaN grating-gate structures, our results have a general character and are applicable to any other semiconductor-based plasmonic crystal structures.
Our work represents the first demonstration of an electrically tunable transition between different phases of THz plasmonic crystals, which is a crucial step towards a deeper understanding of THz plasma physics and the development of all-electrically tunable devices for THz optoelectronics.
\end{abstract}

\maketitle


\section{Introduction}\label{Sec1}

Terahertz (THz) plasmonics~\cite{Otsuji2014, Shur2020} is a newly emerging field of physics and technology studying the wide class of phenomena relating to the peculiarities of interaction of electromagnetic (\textit{em}) waves of THz frequency range with collective oscillations of electron gas (plasmons) in semiconductor micro- and nanodevices. 
Generally, two-dimensional (2D) plasmons can be observed in film-like samples when the scale of electron gas confinement is significantly smaller than the wavelength of the plasmons propagating along the sample. 
This spatial localization of electron gas oscillations imparts unique physical properties to 2D plasmons that fundamentally differ from those of their bulk counterparts. 
In the 2D scenario, one of the most important properties is the ability to tune and manipulate the plasmon parameters through the sample's geometry and an electric field.
The great interest focuses on the quantum wells (QWs) or 2D material-based structures (including graphene structures), where 2D plasmons can be excited in the ultra-thin conductive layer~\cite{Otsuji2014, Shur2020, Huang2017, knap2009field}.
This interest comes from the fact that nanoscale devices using the excitation of the 2D plasma waves are promising for THz optoelectronics detection~\cite{knap2009field, Shaner2005, Knap2011, Kurita2011, Spisser2016, Delgado2022} as well as amplification or generation~\cite{dyakonov1993, Michailov1998, Kachorovskii2012, Mikhailov2016, Petrov2017, Svintsov2019, KorPRB2020}.

The history of 2D plasmons began more than 50 years ago with the publication of a seminal paper by A.~V.~Chaplik~\cite{Chaplik1972} and the subsequent review~\cite{Chaplik1985}. These works pointed out that 2D plasmon frequency $\omega_{p}$ exhibits significant wavevector dispersion, $\omega_{p}({\bf q})$, even in the long-wavelength limit, while 3D plasmons are essentially dispersionless. 
In the retardationless approximation, the electric field associated with the oscillations of 2D plasma waves has two components: longitudinal and transverse. 
Both components penetrate outside the conductive layer, determining the strong dependence of $\omega_{p}({\bf q})$ on the surrounding dielectric or conductive layers.

The first experimental observations of 2D plasmon resonances are related to the structures formed by metallic grating gates integrated with silicon inversion layers~\cite{Allen1977, Tsui1978, Tsui1980}. Later, the similar studies were performed for \text{InGaAs/InP}~\cite{Peale2013}, AlGaAs/GaAs~\cite{Bialek2014}, AlGaN/GaN~\cite{muravjov2010,Qin2016,jakvstas2017,shalygin2019} and, recently, for the graphene structures~\cite{Yan2012,Yan2015,li2019current}.


The field of 2D plasmonics has seen a surge of interest since the landmark work of Dyakonov and Shur~\cite{dyakonov1993}. Their theoretical predictions showed that the steady-state current in the channel of a sub-micrometer-sized field-effect transistor (FET) can excite large-amplitude plasma oscillations, resulting in the generation of THz radiation with frequency controlled by the gate voltage. An advanced version of plasmonic FETs is the grating-gate structure, in which a periodic charge carrier density profile with intercalated high and low-density regions is formed. Such structures are known as plasmonic crystals (PCs). Researchers, including V.~Popov, D.~Fateev, and their colleagues~\cite{Popov2010, Popov2011, Popov1993}, have extensively studied the THz plasmonic absorption properties of grating-gate-based FETs.

For the case of PC structures, there are several theoretical predictions~\cite{watanabe2013gain, takatsuka2012gain, boubanga2012ultrafast} which denote a possibility to create a solid-state analogue of vacuum amplifiers and generators based on the Smith-Purcell effect or related phenomena when 2D electrons beam moving in the periodic potential of the grating can trigger plasmon instability and THz generation. It is expected that semiconductor PC amplifiers and sources can be tuned electrically by the gate voltages (electron density) and/or the lateral current (electron drift velocity).
Also, it has been demonstrated recently that 2D plasmon resonance in PC structures is accompanied by a significant phase shift between an incident and transmitted waves~\cite{Pashnev2020}. This phenomenon has led to the suggestion that the electrical control of this phase shift in grating-gate PC structures~\cite{Kor2022} could be exploited to create a phased array antenna in the THz range.

Very important experimental results~\cite{Knap2020} on graphene PC structures have shown that indeed 2D grating-gate coupled plasmons can open the way towards THz radiation amplifiers.
To interpret these results a simplified analytical model theoretically developed by V. Kachorovski et al. was applied~\cite{Kachorovskii2012}. 
Despite important theoretical background and the first promising experimental works on grating-gate plasmonic THz sources and amplifiers - the 2D plasmon resonances in PC structures are still not fully understood.

Indeed, comprehensive theoretical and experimental analyses of the resonant plasmon properties of such structures are missing in the literature.
This comes from two main reasons: 
(i) experimental/technological difficulties - THz experiments require systems of millimeter size (the wavelength of 1~THz radiation is $\sim0.3$~mm) whereas typical THz oscillating plasmon cavity is of a few hundred nanometers, which means that in THz experiments one needs samples composed of thousands of gates (grating fingers) with negligible leakage currents (zero defects large surface gate isolation) - to reach the regime of an efficient electrical tuning of plasmon resonances; and
(ii) theoretical challenges - the grating gate on 2DEG forms a nonuniform system composed of intercalated gated and ungated plasmon cavities, for which no analytical solution exists. Theoretical predictions/interpretations require advanced numerical calculations.


In this work, we addressed these two key issues and conducted both experimental and theoretical investigations of 2D plasmons in grating-gate PC structures based on AlGaN/GaN heterojunctions as a function of the gate-to-channel voltage. We hereby present: 

(i) technology of large surface (a few square mm) grating-gate structures with thousands of nanometer gates controlling carrier density (more than two decades) and negligible gate leakage currents (below 100~nA/mm$^{2}$); and  

(ii) rigorous electrodynamic approach that enables accurate numerical simulations of the optical characteristics of PCs. This approach has been validated through excellent agreement with experimental data, enabling a better understanding of the basic tuning mechanisms of plasmon resonances in PCs. Furthermore, this approach also allows for a comprehensive discussion of existing phenomenological approximations.

In particular, we have identified two distinct plasmonic crystal phases: a \textit{delocalized} phase observed at a low carrier density modulation degree and a \textit{ localized} phase dominating when the gated part of the 2DEG approaches a total depletion regime (strong modulation). Additionally, we have demonstrated an electrically-controlled continuous phase transition between these two phases. For experimental studies, we have chosen grating-gate structures based on AlGaN/GaN heterojunctions. However, our results have a general character and are applicable to any other semiconductor-based PC structures.

The paper is organized as follows: Section~\ref{Sec2} includes the theoretical model, formalism and mathematical details of a numerically stable and fast-convergence algorithm developed for the solution of Maxwell's equations.  
Preliminary calculations of far-field characteristics, particularly transmission coefficients at different modulation degrees of 2DEG are also presented here. Details of the fabrication, basic characterizations and THz measurements are collected in Section~\ref{Sec3}. 
The main experimental results and their comparison with electrodynamic simulations of particular PC structures are presented in Section~\ref{Sec4}.
Section~\ref{Sec5} is dedicated to detailed analyses of the results based on the calculations of near-field characteristics, particularly, the spatial distribution of \textit{em} field and local absorptivity in different PC phases.
This section is finalized with conclusions.

\section{Electrodynamics of grating-gate PC structures}\label{Sec2}
\subsection{General consideration}

The oscillation frequency of 2D plasma excitation propagated along infinitely-long  and the delta-thin conductive layer is given by the following formula~\cite{Chaplik1985, Popov2011}:

\begin{equation}
\omega_{p}=\sqrt{\frac{2\pi e^2 n |q|}{m^{*}\epsilon_{\text{eff}}(q)}},
\label{2Dplasmons}
\end{equation}
where $e$ is the electron charge and $n$ is the concentration of 2DEG.
Throughout this article, all units in formulas are given in the CGS system of units.
This formula is obtained in the hydrodynamic consideration of electron transport (i.e. as a solution of the eigenvalue problem of the Euler-Poisson system of equations) neglecting viscosity effect~\cite{Rudin2010} and thermal distribution of electrons in momentum space~\cite{Totsuji1976, KorPRB2020}.
Also, Eq.~(\ref{2Dplasmons}) assumes a spatially uniform distribution of steady-state electron concentration in the conductive layer and parabolic electron dispersion law with an effective mass, $m^{*}$. Effective dielectric permittivity, $\epsilon_{\text{eff}}(q)$,  depends on the surrounding of the conductive layer.

\begin{figure*}[t!!!]
\includegraphics[width=1\textwidth]{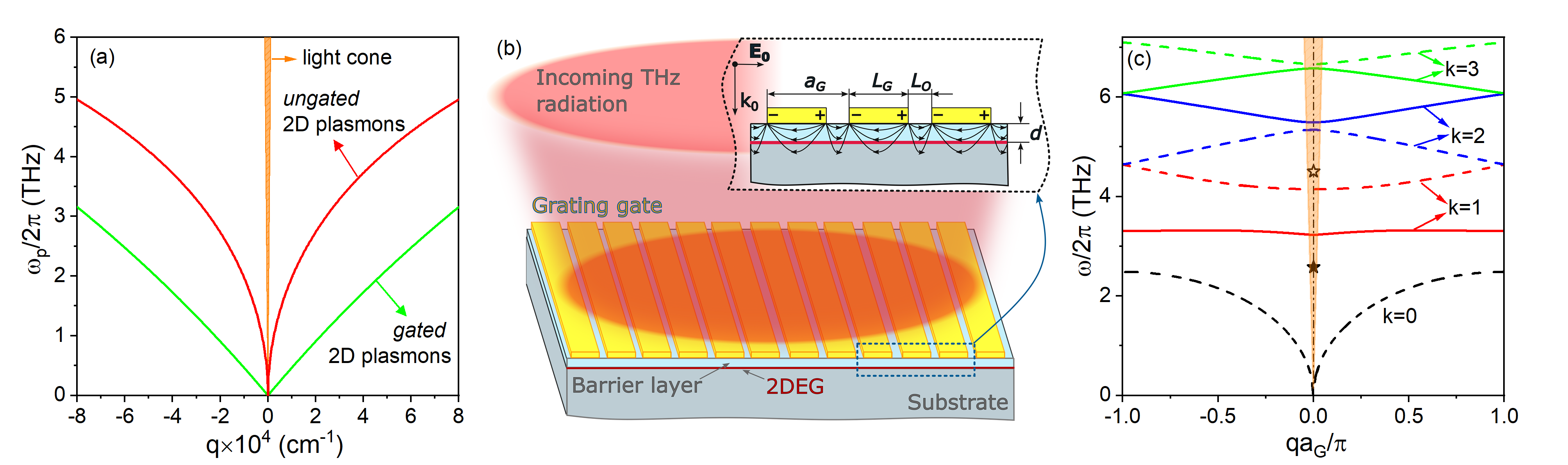}
 \caption{Dispersions of 2D plasmons given by Eq.~(\ref{2Dplasmons}) for ungated (Eq.~(\ref{eps_ungated})) and gated (Eq.~(\ref{eps_gated})) AlGaN/GaN QW heterostructures with the following parameters: $n=10^{13}$ cm$^{-2}$, $d=50$ nm, $m^{*}=0.2\times m_{e}$
 ($m_{e}$ is the free electron mass) and $\epsilon_{bar}=\epsilon_{buf}=8.9$ (a); a sketch of the plasmonic crystal structure (b), and plasmon spectrum under the grating in Brillouin zone-folding representation at $a_{G}=1$ $\mu$m, $f=0.4$ and the same other parameters (c). These results are obtained in the electrostatic limit, using the algorithm of paper~\cite{Popov1993}.
 Stars indicate the resonant frequencies of gated (star) and ungated (open star) plasmons calculated from Eq.~(\ref{2Dplasmons}) at $q=2\pi/a_{G}$.}
  \label{fig1}
\end{figure*}

In literature, the two simplest types of 2D plasmons are mainly discussed. The first type is the ungated 2D plasmons, which can be excited in the conductive layer of the QW heterostructure. For these plasmons,

\begin{equation}
\epsilon_{\text{eff}}(q)=\frac{1}{2}
\left[\epsilon_{\text{buf}}+\epsilon_{\text{bar}}\frac{1+\epsilon_{\text{bar}}\tanh(|q|d)}{\epsilon_{\text{bar}}+\tanh(|q|d)} \right],
\label{eps_ungated}
\end{equation}
where $\epsilon_{\text{buf}}$ and $\epsilon_{\text{bar}}$ are dielectric constants of a buffer and barrier layers, respectively, and $d$ is the thickness of the barrier layer. At this, the thickness of the buffer layer is assumed to be much larger than the plasmon wavelength.
The second type is the gated 2D plasmons. These plasmons can be excited in the screened QW heterostructure where the barrier layer is covered by a perfect metallic gate. For them,

\begin{equation}
\epsilon_{\text{eff}}(q)=\frac{1}{2}
\left[\epsilon_{\text{buf}}+\epsilon_{\text{bar}}\coth(|q|d)\right].
\label{eps_gated}
\end{equation}

If $|q|d<<1$, then $\epsilon_{\text{eff}}(q)\approx(\epsilon_{\text{buf}}+1)/2$ for ungated plasmons and $\epsilon_{\text{eff}}(q)\approx\epsilon_{\text{bar}}/2|q|d$ for gated ones. As a result, in this limit, ungated plasmons have square-root dispersion, $\omega_{O}(q)$,  and gated plasmons are described by the linear dispersion, $\omega_{G}(q)$:

\begin{equation}
\omega_{O}=\sqrt{\frac{4\pi e^2 n}{m^{*}(\epsilon_{{\text{buf}}}+1)}|q|},\,\,
\omega_{G}=|q|\sqrt{\frac{4\pi e^2 n d}{m^{*}\epsilon_{\text{bar}}}}
\label{limits}
\end{equation}

Remarkably, for typical parameters of QW heterostructures, characteristic frequencies of both types of 2D plasmons belong to the THz frequency range at the micron and sub-micron scale of their wavelengths, $\lambda_p = 2 \pi / |q|$.
The latter is illustrated in Fig.~\ref{fig1}~(a). Also, it is seen that the gated plasmons have smaller phase velocity than ungated plasmons.

 It should be noted that 2D plasmons, as well as another class of surface \textit{em} waves, such as surface plasmon polaritons~\cite{Baltar12} or surface phonon polaritons~\cite{Huber2008}, cannot be directly excited by incident \textit{em} radiation. 
 The 2D plasmon-photon interaction in the laterally-uniform sample is forbidden because the wavevector of 2D plasmons is much greater than the wavevector of the incident radiation (see the light cone in Fig.~\ref{fig1}~(a)) at a given frequency, i.e., it is impossible to satisfy energy and momentum conservation laws simultaneously.

 In order to provide an effective coupling between the 2D plasmons and \textit{em} waves, the electron conductive channel should be supplemented by a lateral microstructure of the sample (plasmonic structure), for example, by using subwavelength metallic grating (Fig.~\ref{fig1}~(b)).
 Such grating, formed by long metallic strips, plays the role of a broadband antenna that can effectively couple incident radiation with plasmon oscillations of 2DEG, as well as a polarizer for incident THz radiation and concentrator of \textit{em} energy in near-field~\cite{Kor2012, Kor2014}.

In the framework of the simple physical interpretation, the incoming THz wave with p-polarization (electric field is oriented perpendicularly to the grating strips) induces the instantaneous dipoles (electrical vibrators) at the edges of the strips. In turn, these dipoles excite the particular 2D plasmons. The efficiency of such coupling and selection of the resonant plasmonic mode is controlled by the grating geometry: grating period, $a_{G}$, grating filling factor, $f=L_{G}/a_{G}$, as well as by the thickness of the barrier layer, $d$.

On the other hand, the interaction of the incident wave with metallic grating stipulates the formation of the strongly non-uniform distribution of the \textit{em} field (near-field) in the proximity of the grating coupler. For the actual case of subwavelength grating, the near-field is composed of $k$ evanescent ($k = 1,2, 3...$) waves and one propagating ($k=0$) wave. 
Evanescent waves decay exponentially away from the grating plane $\sim \exp(-2 \pi k z /a_G)$ with the periodicity in the lateral direction. 
The evanescent waves with in-plane wavevector $q = 2 \pi k / a_G$  excite plasmons of different $k$-orders when the frequency of the incident THz wave coincides with the frequency of a particular eigen plasmon mode of 2DEG under the grating. 
As a result, part of \textit{em} energy transfers to the electron subsystem and resonant absorption of incident radiation occurs.

More rigorously, the presence of the metallic grating leads to the "folding" and splitting of plasmon dispersion branches (see Fig.~\ref{fig1}~(c)). As seen, in Brillouin zone-folding representation~\cite{Ager1991, Petrov2017}, the 2D plasmon spectrum consists of a set of branches. Therefore, such a plasmonic structure can be called a plasmonic crystal (PC). At small $q$, the branches marked by solid lines in Fig.~\ref{fig1}~(c) become optically active and they can interact with \textit{em} waves. The dashed lines represent the \textit{ dark} plasmon modes. Due to the symmetry of the electric field oscillations, they do not interact with \textit{em} waves~\cite{Popov1993}.

\subsection{Electrodynamic simulations}

The response of the grating-gate PC structures to THz radiation is analyzed by the rigorous electrodynamic simulations based on numerical solutions of Maxwell’s equations in the framework of the integral equations (IEs) method following the pioneering work of  Ref.~\cite{Popov2010}. Generally, this technique uses Green function formalism and is based on a reduction of Maxwell’s system of equations to the linear IEs in the coordinate space. The latter can be solved, for example, using Galerkin schemes. In contrast to well-known and widely-applied Fourier-modal methods~\cite{Gaylord, KorRCWA2021},
the IE technique provides much faster and guaranteed convergence of the results with a given accuracy.
\begin{figure}[t!!!]
\includegraphics[width=0.45\textwidth]{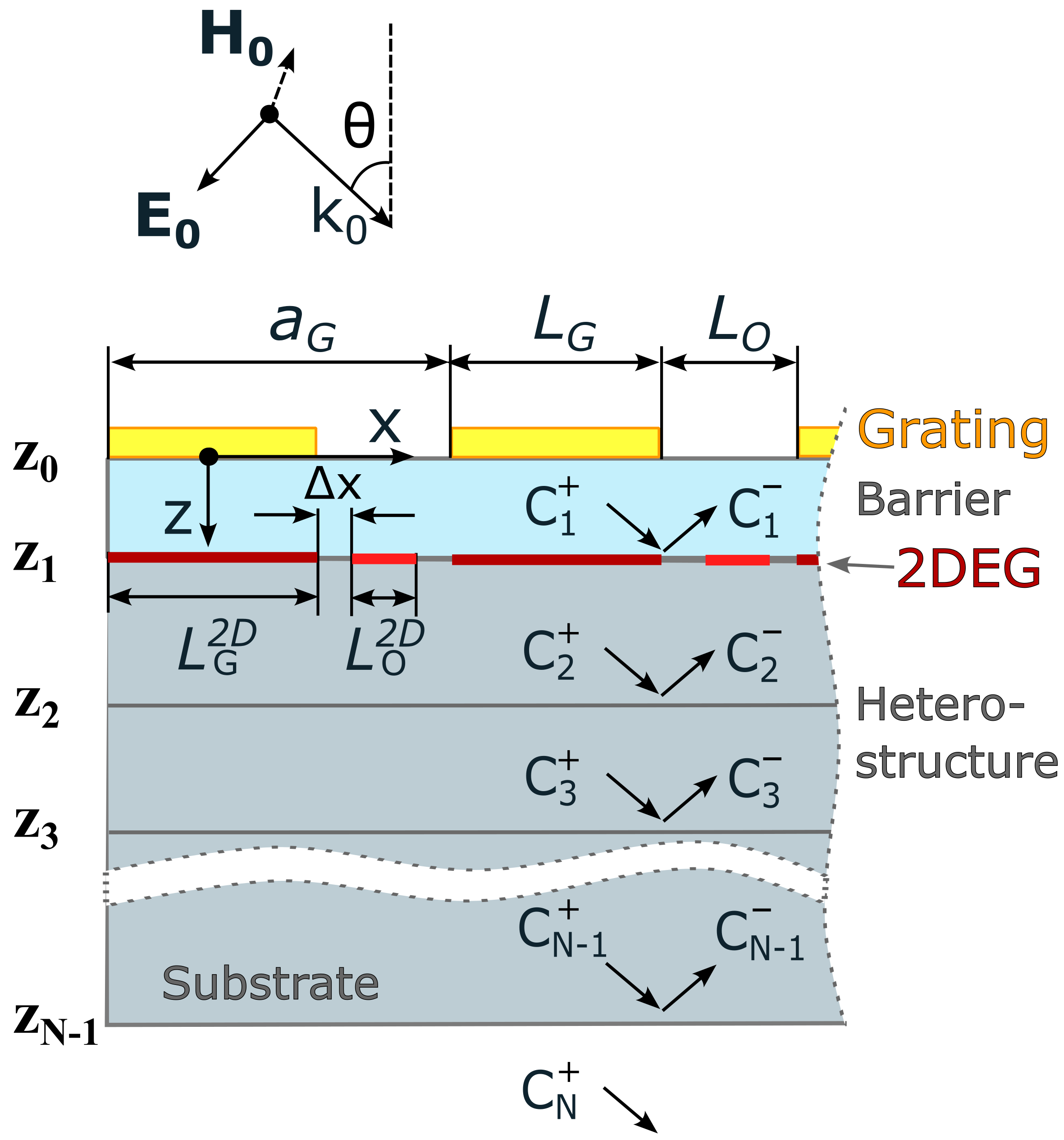}
\caption{Sketch of the considered plasmonic structure.}
\label{Fig2}
\end{figure}

In general, the PC structure can be considered as a multi-layer semiconductor heterostructure with 2DEG conducting channel integrated with metal grating. The geometry of such a structure is sketched in Fig.~\ref{Fig2}.
In this work, the IE method of Ref.~\cite{Popov2010}, is generalized to be valid for arbitrary geometries of the metal grating and the semiconductor heterostructures.
This modification takes into account different forms of the spatial profiles of 2DEG conductivity and the possibility of incline incidence of electromagnetic waves.  As it is shown in Fig.~\ref{Fig2}, the considered PC structure consists of the $N-1$ layers placed between $N$ interfaces. Each layer is described by dielectric permittivity, $\varepsilon_{r=1..N-1}$, (where subscript $r$
numerates here different layers of the structure). The whole structure is placed between two dielectric half-spaces with $\varepsilon_{0}$ and $\varepsilon_{N}$.  The periodic metallic grating is deposited on the top of the first layer ($z=z_0$). The grating is characterized by the grating period, $a_G$, the gate finger width, $L_G$, and the grating opening width, $L_O$. Such grating is formed by infinitely long (along the y-axis) parallel strips of the metal with 2D conductivity, $\sigma_{M}$. The spatial profile of the grating conductivity $\sigma^{G}(x)$ is taken as follows:
\begin{equation}\label{sigmaG}
\sigma^{G}(x)\!=\!\left\{\!\!
\begin{array}{ll}
\sigma_{M}v_G(x), &-\frac{L_{G}}{2}\!<\!x\!<\!\frac{L_{G}}{2}\\\\
0,&\frac{L_{G}}{2}\!<\!x\!<\!a_{G}\!-\frac{L_{G}}{2},
\end{array}
\right.
\end{equation}
where both, metallic grating and 2DEG are assumed as delta thin. The 2DEG conductive layer has the coordinate $z=z_1$ and it is described by the 2D conductivity, $\sigma^{2D}(x)$ with the following spatial profile:
\begin{equation}\label{sigma2D}
\sigma^{2D}(x)\!=\!\left\{\!\!
\begin{array}{ll}
\sigma^{2D}_{G}v_G^{2D}(x), &-\frac{L_{G}^{2D}}{2}\!<\!x\!<\!\frac{L_{G}^{2D}}{2}\\\\
\sigma^{2D}_{O}v_{O}^{2D}(x), &\frac{L^{2D}_{G}}{2}\!+\!\Delta x\!<\!x\!<\!a_{G}\!-\frac{L^{2D}_{G}}{2}-\!\Delta x
\end{array}
\right.
\end{equation}
where $\sigma^{2D}_{G,O}=e\mu_{G,O}n_{G,O}$, and $v_G(x),\,v_G^{2D}(x),\,v_{O}^{2D}(x)$ are dimensionless conductivity profiles of the metallic grating strips, gated and ungated regions of the 2DEG, respectively.
Under applied gate voltage, $V_G$, the concentrations $n_G$ and $n_O$ can be essentially different. Also, we introduced the parameter, $\Delta x$, which describes the effective width of the ungated region of 2DEG,  $L_{O}^{2D}=a_{G}-L^{2D}_{G}-2\Delta x$ (note, the method in Ref.~\cite{Popov2010} was developed only for the case of $\Delta x = 0$). This approach accounts for the fact that the depletion region under the gate extends laterally beyond the geometrical gate finger dimension.

We suppose that the whole structure is uniformly illuminated by the plane \textit{em} wave at the incidence angle, $\theta$, having TM-polarization, i.e, $x$-component of the electric field $E_{0,x}$ orients perpendicularly to the grating strips. At this, the electric field of the \textit{em} wave induces in the grating strips instantaneous dipoles (see, for example, Ref.~\cite{Kor2014}) which form a spatially inhomogeneous field in the near-field zone. This one can effectively excite the plasmon oscillations in 2DEG.

In such geometry, the non-zero components of the \textit{em} wave are $E_x$, $E_z$, $H_y$. The master equation for the electric components, $E(\textbf{r}, t)$, of the \textit{em} field which is a result of the interaction of incident wave with plasmonic structure is:
\begin{align} \label{Maxwell}
  \textrm{rot}\;\textrm{rot}{\bf E}(\textbf{r},t)+\frac{\epsilon({\textbf{r}})}{c^2}\frac{\partial^{2} {\bf E}(\textbf{r},t)}{\partial t^{2}}=-\frac{4\pi}{c^2}\frac{\partial {\bf j}(\textbf{r},t)}{\partial t}.
\end{align}

Here, the term of the displacement current contains dielectric permittivity $\epsilon(\textbf{r})$ which is a step-like function and constant within each medium. Conduction current includes the current density in the metallic grating, $\textbf{j}^G (\textbf{r}, t)$, and in the 2DEG conduction layer, $\textbf{j}^{2D}(\textbf{r}, t)$. In the delta-thin treatment of both currents,
 \begin{align} \label{Current}
  j_{x}(\textbf{r},t)=j^{G}(x,t)\delta(z-z_{0})+j^{2D}(x,t)\delta(z-z_{1}).
\end{align}
where $\delta$ denotes the Dirac delta function. Because of the periodicity of the plasmonic structure, we can introduce the spatial Fourier transform of the electric field and the currents with respect to the $x$-coordinate:
\begin{equation}
 \left[
\begin{array}{c}
E_{\{x,z\}}(\textbf{r},t) \\
 j^{G}(x,t)\\
 j^{2D}(x,t)
  \end{array}
   \right]
  \!\! =\!\! \sum^{+\infty}_{k=-\infty}\!
 \left[
 \begin{array}{c}
  E_{\{x,z\},\omega,k}(z) \\
j^{G}_{\omega,k}\\
j^{2D}_{\omega,k}
   \end{array}
   \right]
   \!\exp(i[Q_k x-\omega t]),
\label{FourierTrans}
\end{equation}
In Eq.~(\ref{FourierTrans}) we assumed harmonic temporal dependencies of \textit{em} wave with angular frequency, $\omega$. Periodicity parameter, $Q_{k}=q_{k}+\text{k}_{0}\sqrt{\epsilon_{0}}\sin{\theta}$ ($q_{k}=2\pi k/a_{G}$ and $\text{k}_{0}=\omega/c$). Now, the wave equation (\ref{Maxwell}) can be transformed to the system of the differential equations with respect to the z-coordinate for each $k$-spatial Fourier harmonics of the $x$-component of the electric field in every $r-$th medium ($r=0..N$). So that for $E_{x,\omega,k}$ we obtain:
 \begin{align}\label{Maxwell1}
    \frac{d^2 E^{(r)}_{x,\omega,k}}{dz^2}-\! \lambda^2_{r,k}E^{(r)}_{x,\omega,k}\!=\frac{4\pi i\lambda_{r,k}^2}{\epsilon_{r}\omega}\!\times\nonumber\\
    \left[j^{G}_{\omega,k}\delta(z-z_{0})\!+\!
     j^{2D}_{\omega,k}\delta(z\!-\!z_{1})\right],
\end{align}
where the characteristic parameter
\begin{equation}
     \lambda_{r,k}=\left\{\begin{array}{ll}
\,\,\, \sqrt{Q_k^2-\epsilon_{r}\text{k}_{0}^2}\,, &\ Q_k>Re[\sqrt{\epsilon_{r}}\text{k}_{0}]\,\\
  -i\sqrt{\epsilon_{r}\text{k}_{0}^2-Q_k^2}\,, & Q_k<Re[\sqrt{\epsilon_{r}}\text{k}_{0}]\,,
  \end{array}\right.
  \label{Lambda}
\end{equation}
describes the spatial localization of $k$-spatial Fourier harmonic. In Eq.~(\ref{Lambda}), the upper line corresponds to the evanescent modes and the lower line corresponds to the propagated modes. For the actual case of deeply sub-wavelength grating,  optical characteristics of the structure, such as transmission/reflection coefficients are formed by the single propagated mode with $k=0$.

The formal solutions of system (\ref{Maxwell1}) can be presented as superposition of transmitted ('$+$') and reflected('$-$') waves and one reads as:
\begin{equation}
\label{sol}
E^{(r)}_{x,\omega,k}=\left\{
\begin{array}{l}
C^{+}_{0,k}\exp(-\lambda_{0,k}z)+C^{-}_{0,k}\exp(\lambda_{0,k}z),  z<z_{0} \\
\vdots\\
C^{+}_{r,k}\exp(-\lambda_{r,k}(z-z_{r-1}))+\\
+C^{-}_{r,k}\exp(\lambda_{r,k}(z-z_{r})),  \quad\quad z_{r-1}\leq z\leq z_{r}\\
\vdots\\
C^{+}_{N,k}\exp(-\lambda_{N,k}(z-z_{N-1})),  z\geq z_{N-1}.
\end{array}
\right.
\end{equation}
This system should be supplemented by the following boundary conditions at $r-$th interface:
\begin{align}
\label{Bound1} E^{(r)}_{x,\omega,k}|_{z=z_{r}}=E^{(r+1)}_{x,\omega,k}|_{z=z_{r}},\\
\frac{\epsilon_{r+1}}{\lambda^{2}_{r+1,k}}\frac{dE^{(r+1)}_{x,\omega,k}}{dz}\Big|_{z=z_{r}}\!-\!\frac{\epsilon_{r}}{\lambda^{2}_{r,k}}\frac{d E^{(r)}_{x,\omega,k}}{dz}\Big|_{z=z_{r}}\!=\nonumber \\
=\frac{4\pi i}{\omega}[j^{G}_{\omega,k}\delta_{r,0}+j^{2D}_{\omega,k}\delta_{r,1}].
  \label{Bound2}
\end{align}
Here, $\delta_{r,r'}$ being the Kronecker delta-symbol.
Eq.~(\ref{Bound1}) corresponds to the continuity of the tangential components of the electric fields and the latter Eq.~(\ref{Bound2}) takes into account the discontinuity of the tangential component of the magnetic field due to the presence of the delta-thin conductive layers.
Without loss of generality, we can put that $C^{+}_{0,k}=\delta_{k,0}E_{0,x}$ where $E_{0,x}=E_{0}\cos{\theta}$ and $E_{0}$ is the electric field amplitude of the incident wave.

After some algebraic transformations, we can come to the relationships between Fourier harmonics of the $x$-component of the electric fields in the plane of the metallic grating ($z=z_0$) and 2DEG conductive layer ($z=z_1$), and Fourier-harmonics of the corresponding currents:
\begin{align}\label{Fur_Bas}
E^{G}_{\omega,k}=-\frac{2\pi i}{c\sqrt{\epsilon_{0}}}[Z^{(11)}_{\omega,k}j^{G}_{\omega,k}+Z^{(12)}_{\omega,k}j^{2D}_{\omega,k}]+\eta^{(1)}_{\omega,k}\delta_{k,0}E_{0,x}\nonumber\\
E^{2D}_{\omega,k}=-\frac{2\pi i}{c\sqrt{\epsilon_{0}}}[Z^{(21)}_{\omega,k}j^{G}_{\omega,k}+Z^{(22)}_{\omega,k}j^{2D}_{\omega,k}]+\eta^{(2)}_{\omega,k}\delta_{k,0}E_{0,x}
\end{align}

Here $E^{G,2D}_{\omega,k}$ are $E^{(1)}_{x,\omega,k}(z_{0})$ and $E^{(2)}_{x,\omega,k}(z_{1})$, respectively. The explicit form of the introduced parameters is specified in Supplemental Material~\cite{suppl}.

In the frameworks of the IE method, the system~(\ref{Fur_Bas}) should be transformed into coordinate space and formulated as a system of 3 linear integral equations with respect to currents $j^G (x)$, $j_G^{2D} (x)$, $j_O^{2D} (x)$  induced by \textit{em} wave in the grating strip, gated and ungated regions of 2DEG, respectively. For this, it is necessary to multiply both sides of the system~(\ref{Fur_Bas}) by $\exp(i Q_k x)$, then sum them overall $k$ taking into account local approximation for the currents,
 \begin{align} \label{Current1}
  j^{G, 2D}(x) = \sigma^{G, 2D}(x)E^{G, 2D}(x).
\end{align}


The mathematical details of this procedure, including the method of solution for the obtained IEs and the final formulas used for calculating the transmission, reflection, and absorption spectra (including near-field patterns), are provided in Supplementary Material~\cite{suppl}. In Supplementary Material~\cite{suppl}, we also discuss the convergence of the developed method and compare it with another purely numerical scheme based on finite-element calculations. We utilized the described numerical scheme to investigate the main peculiarities of the THz transmission spectra of the PC structures described below, at different applied gate voltages.

\subsection{Gate voltage tuned plasmonic spectra}
In order to describe the different phases of the plasmonic crystals, we introduce the dimensionless parameter that characterises the modulation degree, $\rho$. It is defined as $\rho(V_G )=(n_O-n_G (V_G))/(n_{O}+n_{G}(V_G)) $, where $n_{O}$ and $n_{G}$ are concentrations of 2DEG in ungated and gated regions of 2DEG, respectively. The $n_G (V_G)$ can be controlled by the applied gate voltage $V_G$.
For the case of the uniform spatial profile,  $n_O = n_G$,  and  $\rho=0$.  For the case when $|V_G|$ is larger than the threshold voltage $|V_{th}|$, $\rho \sim 1$. In the latter case, the space between the drain and source is divided into highly conductive ungated regions and almost depleted gated regions. As shown below, the resonant peculiarities of the transmission spectra strongly depend on modulation degree $\rho (V_G)$.

First, we imply that conductivity profiles of the metallic grating and 2DEG have an abrupt step-like form, i.e., $v_G (x)= v_G^{2D} (x) = v_O^{2D} (x) = 1$ and width of the gated region of 2DEG coincides with grating-finger width, $L_G^{2D} = L_G$.

In calculations, we used the parameters of the typical GaN-based high electron mobility heterostructure used in the experiment (see Section~\ref{Sec3}).
The whole structure consists of 4 bulk-like dielectric layers:  AlGaN barrier layer with thickness $d_{1}=23$ nm, GaN buffer layer with $d_{2}=255$ nm placed directly on AlN nuclear layer with $d_{3}=62$ nm (see Table I  in Section~\ref{Sec3}). In order to avoid the emergence of the Fabry-P$\acute{\text{e}}$rot fringes, here, we consider a membrane-like SiC substrate with a thickness  $d_{4}=1$ $\mu$m.

For the 2DEG densities we assumed the linear dependence of $n_{G}(V_G)$ provided by so-called “plane capacitor” approximation:
\begin{equation}
n_{G}=\left\{
\begin{array}{cc}
 \displaystyle{\frac{\varepsilon_{1}V_0}{4\pi e d_{1}}},    & \text{at}\,\, V_0\geq 0   \\\\
     0, &  \text{at}\,\, V_0<0,
\end{array}
\right.
\label{capacitor}
\end{equation}
where $V_0 = V_G - V_{th}$ is the gate voltage swing, $V_{th}$ is the threshold voltage of channel depletion. At this, we set that $n_O$ is a constant and equal to $9 \times 10^{12}$ cm$^{-2}$, threshold voltage $V_{th}=-2.9 \pm 0.3$ V and $\varepsilon_{1} = 8.9$ (see Section~\ref{Sec3}). 

The metallic grating is assumed to be gold and delta-thin with 1.5 $\mu$m period, grating filling factor, $f=0.6$, and dispersionless 2D conductivity $\sigma^{M} = 2\times10^{12}$ cm/s. It corresponds to the bulk conductivity of $4\times10^{17}$ s $^{-1}$ and the grating thickness of $50$ nm.

High-frequency mobilities of 2DEG in the gated and ungated region are assumed to be equal, $\mu_G = \mu_O$, and both obey to Drude-Lorentz model, $\mu_{G,O} = \mu/(1 - i 2\pi\nu\tau)$  with frequency, $\nu$, and effective scattering time, $\tau=\mu m^*/e$. Here, we set $\mu = 7500$ cm$^2$/Vs that corresponds to the scattering time $\tau = 0.85$ ps at electron effective mass in GaN, $m^* = 0.2\times m_e$.
All calculations are performed at normal incidence, $\theta=0$.

\begin{figure}[t!]
\includegraphics[width=0.45\textwidth]{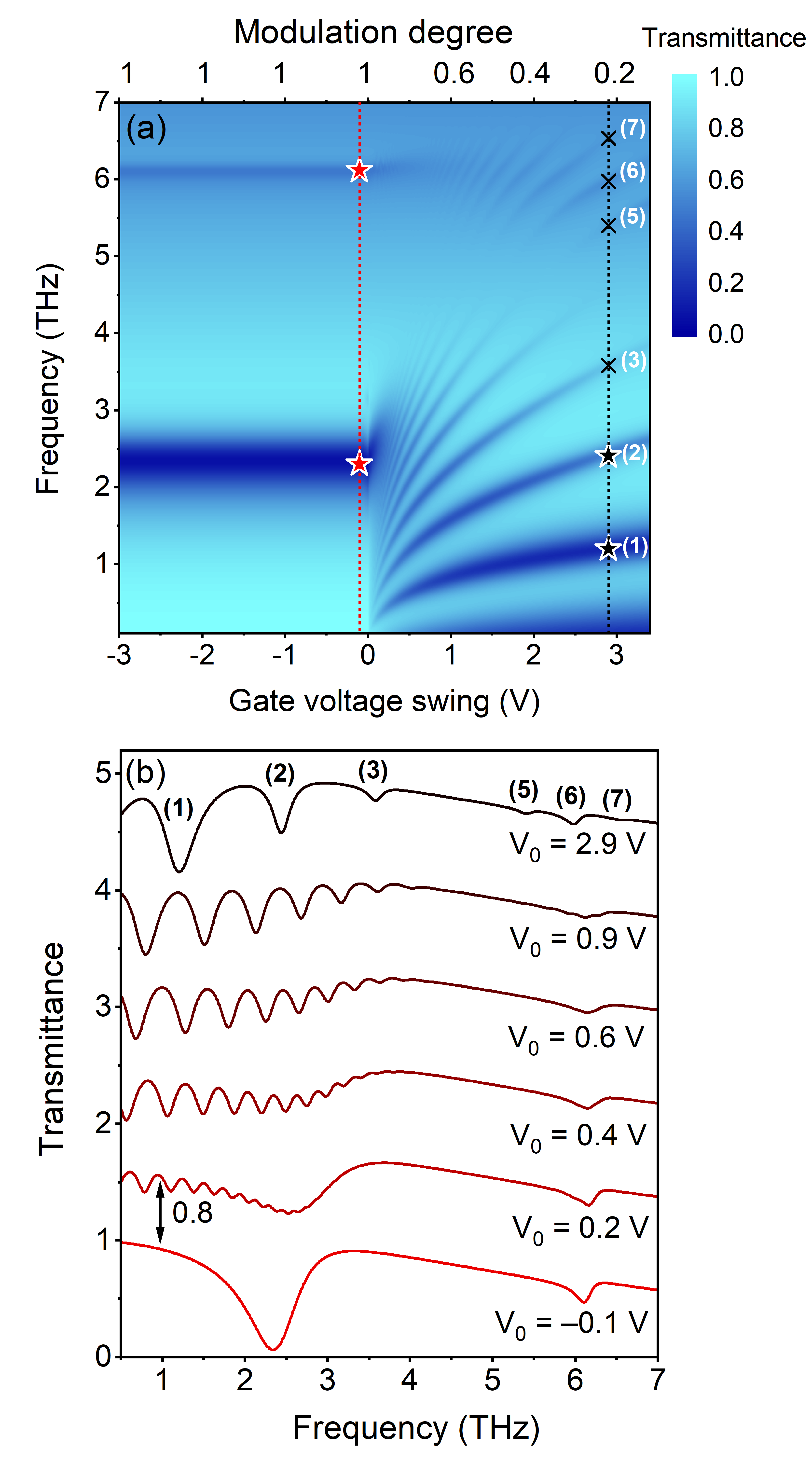}
\caption{Calculated contour plot of the transmittance in the plane ${V_0-\nu}$ (a). 
Transmittance as a function of frequency at the selected values of  $V_0$ (b). 
Vertical dashed lines in panel~(a) show the gate voltage swing values 2.9~V and -0.1~V for plots in panel~(b).
The fundamental resonance "(1)" and visible high-order resonances "(2)", "(3)", "(5)", "(6)", and "(7)" are marked at $V_0 = 2.9$~V.
Stars in panel~(a) mark the points for the near-field simulations to be discussed in section~\ref{Sec5}.}
\label{Fig3}
\end{figure}

Fig.~\ref{Fig3}~(a) demonstrates the calculated contour plot of the transmittance in the plane “gate voltage swing - frequency” in the wide frequency range of 0.1 - 7~THz. One can clearly see two different regions (below and above the threshold voltage) that we attribute to two different plasmonic crystal phases. In the first one, above-threshold voltage, $V_0 > 0$, the transmission spectra possess multiple resonances that are tuned by the gate voltage and can be attributed to the resonant excitation of 2D plasmons of different orders in 2DEG with a spatially modulated profile of electron concentration. We can identify in the frequency range of 0.1 - 4~THz the most intensive 1-st order and less intensive 2-nd, and 3-rd order plasmon resonances (see transmittance spectra in Fig.~\ref{Fig3}~(b) for $V_0 = 2.9$~V). 

With the decrease of $V_0$ (an increase in modulation degree), these resonances are red-shifted and the resonances of a higher order appear to show even up to six resonances as it is seen in the transmittance spectra for $V_0 = 0.9$~V in Fig.~\ref{Fig3}~(b). At $V_0 = 0.6$~V (see Fig.~\ref{Fig3}~(b))  the transmittance exhibits a multi-resonant structure and we can say about the formation of the quasi-continuous plasmonic bands.  

With even further gate voltage swing decrease one can observe the gradual transition to the second plasmonic crystal phase with gate voltage independent modes.  
This transition is clearly seen in the swing voltage range of $0.4$-$0$~V (see curves in Fig.~\ref{Fig3}~(b) calculated for $V_{0}=0.4$ V and $V_{0}=0.2$ V), where the specific evolution of THz transmission spectra with initialization of the formation of single resonant mode regime in the frequency range of 0.1 - 4~THz occurs.
At negative swing voltages, the carrier density under the gated regions is negligible and, therefore, this mode corresponds to the resonant excitation of the plasma oscillations in the limiting case of electrically-formed 2DEG-strip grating~\cite{Michailov1998, Schaich2000, Nosich2017}.


As we will show later, while the phase of PC above the threshold voltage ($V_0>0$) is characterized by \textit{em} energy absorption both, in the gated and ungated regions, in the phase at $V_0<0$ \textit{em} energy is absorbed only in ungated regions. Therefore, we call these phases \textit{delocalized} and \textit{localized} ones, respectively. 

The contour plot in Fig.~\ref{Fig3}~(a) illustrates the phase transition scenario between different PC phases. When the applied voltage exceeds the threshold value ($V_0 > 0$), PC exhibits a well-developed \textit{delocalized} phase. In this phase, multiple resonances appear in the transmission spectra, corresponding to various orders of 2D plasmon excitations.
As the voltage approaches $V_0 \sim 0$, the transition from the \textit{delocalized} phase to the \textit{localized} phase of PC begins. During this transition, the multi-resonant structure in the transmission spectra disappears, particularly within the frequency range of 4~THz to 5~THz, where none of the plasmon resonances exists.
The higher-order resonances in the \textit{delocalized} phase that are resonant with \textit{localized} plasmons start to amplify. This amplification is clearly observed around 2.4~THz and 6.2~THz. Eventually, these higher-order resonances transform into strong plasmon resonances in the \textit{localized} phase of PC.

\section{EXPERIMENTAL DETAILS}\label{Sec3}
\subsection{Fabrication of AlGaN/GaN Grating-Gate PC Structures}

AlGaN/GaN high electron mobility material system was chosen for the experimental studies. This kind of heterostructure provides high electron density (up to $10^{13}$ cm$^{-2}$) on the AlGaN/GaN interface, which is difficult to achieve in any other 2D system. 
Thanks to that plasma oscillations in 2D GaN-based structures of micrometer size can easily reach THz frequencies.

\begin{figure}[t!!!]
\includegraphics[width=0.45\textwidth]{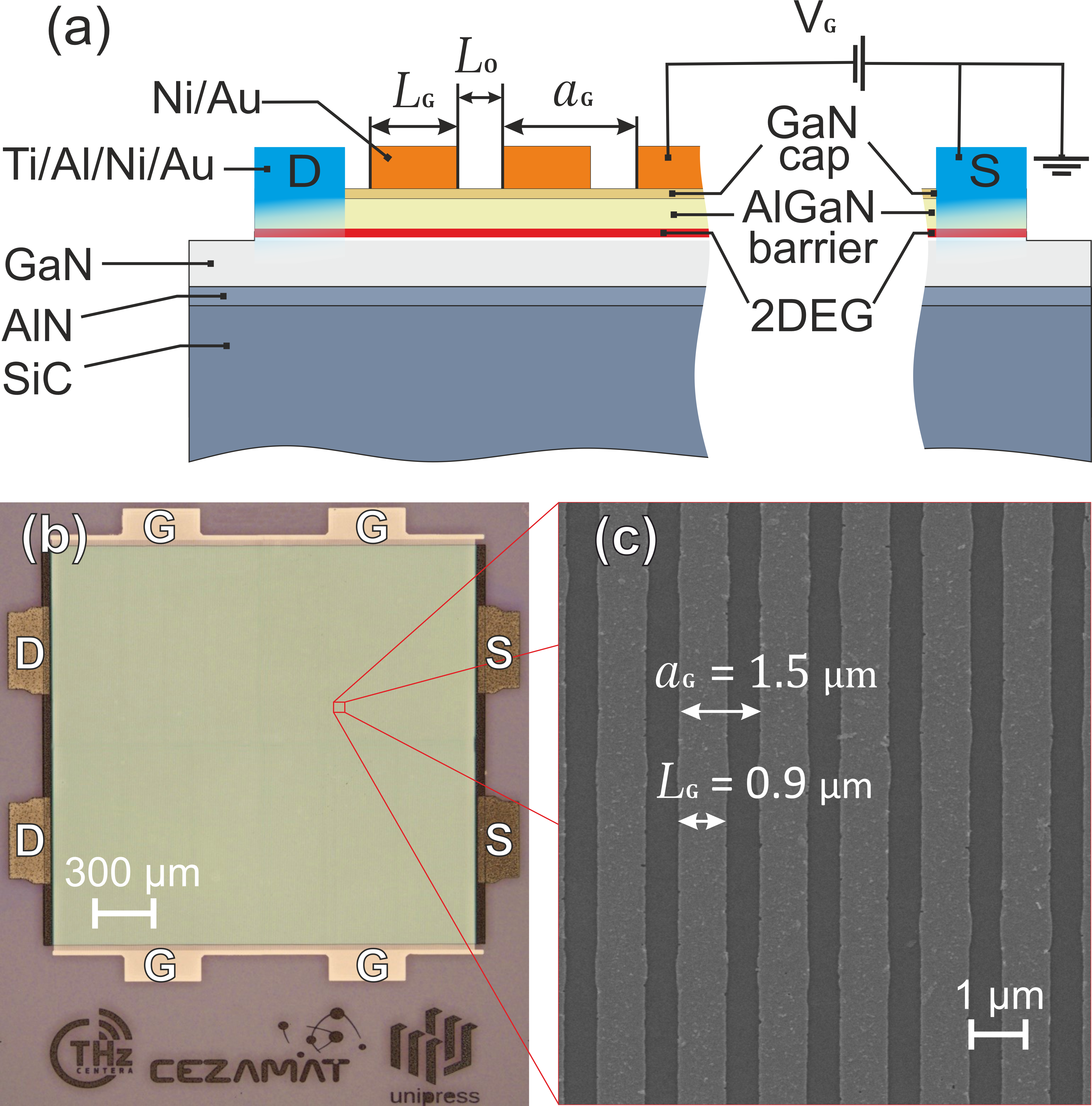}
\caption{\label{Fig4} Cross-section schematic view of the investigated PC structures (a), where S and D are source and drain terminals; optical microscope photo of one of the fabricated grating-gate PC structure (b); SEM image of a grating segment with $a_G=1.5~\mu m$, $L_{G}=0.9~\mu m$, and $L_{O}=0.6~\mu m$ (c).}
\end{figure}

AlGaN/GaN heterostructures were grown by the metalorganic vapor phase epitaxy (MOVPE) method on a 4-inch diameter, 500 $\mu$m-thick semi-insulating SiC substrate. The semiconductor stack consists of a 2.4 nm GaN cap, a 20.5 nm Al$_{0.25}$Ga$_{0.75}$N barrier, and a 255 nm GaN buffer grown directly on 62 nm-thick AlN nuclear layer on SiC substrate. Heterostructure parameters including the permittivity of each layer are listed in Table I. During the fabrication of the plasmonic structures (schematically illustrated in Fig.~\ref{Fig4}~(a)), we paid special attention to the quality of Schottky barrier contact (grating-gate coupler) and ohmic (source and drain) contacts to the 2DEG channel.

The ohmic source and drain contacts (marked as S – source and D – drain in Fig.~\ref{Fig4}) were formed by thermal evaporation of Ti/Al/Ni/Au (150/1000/400/500 \AA) metal stack and following rapid thermal annealing at 800 \textdegree C in a nitrogen atmosphere for 60 s. Such procedure allows achieving the contact resistance of $0.75 \pm 0.08$ Ohm$\times$mm, which was measured using the transmission line method on 30 test structures.

\begin{table}[b!]
\label{tab:structure}
\caption{AlGaN/GaN heterostructure parameters.}
\centering
\renewcommand\arraystretch{1.2}
\begin{ruledtabular}
\begin{tabular}{|| >{\raggedright\arraybackslash}p{0.33\linewidth} | >{\centering\arraybackslash}p{0.25\linewidth} | >{\centering\arraybackslash}p{0.25\linewidth}  | >{\centering\arraybackslash}p{0.08\linewidth} ||}
Layer from the top &	Thickness & Permittivity & Ref.\\
 \hline\hline
GaN cap layer& 2.4~nm & 8.9 &\cite{bougrov2001properties}\\
\hline
Al$_{0.25}$Ga$_{0.75}$N & 20.5~nm & 8.9 &\cite{bougrov2001properties}\\
\hline
GaN	& 255~nm & 8.9 &\cite{bougrov2001properties}\\
\hline
AlN & 62~nm & 8.5& \cite{bougrov2001properties}\\
\hline
SiC	& 500~$\mu$m & 9.7 &\cite{Tarekegne2019}\\
\end{tabular}
\end{ruledtabular}
\end{table}

In order to achieve the low leakage current in high-area grating-gate structures, a special procedure was used to fabricate the Schottky contacts. It started with chemical cleaning of the AlGaN surface. It included the following steps: removing the possible organic contaminants by immersing the samples into an ultrasonic bath filled with dimethyl sulfoxide for 5 minutes followed by cleaning in acetone and isopropanol. Then the natural surface oxides were removed by immersing the samples into 10\% hydrofluoric acid for 5 minutes and 37\% hydrochloric acid for 10 minutes. The chemical cleaning was finished by rinsing in deionized water and drying with an N$_2$ gun.

To avoid any oxidation, after cleaning the samples were immediately placed in the vacuum chamber of an electron beam lithography (EBL) system.  EBL patterning was performed on large $1.7 \times 1.7$ mm$^2$ active areas defining the grating gates. The exposed and developed samples were precisely cleaned from any photoresist residue using oxygen plasma in an inductively coupled plasma-reactive ion etching (ICP-RIE) system. Finally, the grating gates were formed by thermal evaporation of Ni/Au (150/350~\AA) and metal lift-off. The described procedure allowed us to reach the gate leakage current density as low as $J_{GS} \approx 10^{-6}$ A/cm$^2$, which was sufficient to effectively control the 2DEG density under the large area (> 1~mm$^2$) metal gratings.

Three samples with different geometry of the grating-gate coupler were fabricated and investigated. A schematic cross-section of investigated plasmonic structures is shown in Fig.~\ref{Fig4}~(a). Optical microscope and scanning electron microscope (SEM) images of one of the studied devices are shown in Fig.~\ref{Fig4}~(b) and Fig.~\ref{Fig4}~(c), respectively. 
As can be seen in Fig.~\ref{Fig4}~(b), all grating metal fingers were connected together. All structure terminals were provided with special pads for the bonding wires marked in Fig.~\ref{Fig4}~(b) as "G", "S", and "D" letters.
The 2D electron concentration $n_G$ under the grating-gate electrode was controlled by the gate voltage, $V_G$, referenced to the source and drain electrodes connected together. The geometrical parameters of the grating-gate coupler of the three investigated samples labelled as 13s, 7s and 8s are listed in Table II.
\begin{table}[hbt]
\label{tab:grating}
\caption{Grating-gate coupler parameters of investigated PC structures.}
\centering
\renewcommand\arraystretch{1.3}
\begin{ruledtabular}
\begin{tabular}{|| >{\raggedright\arraybackslash}p{0.4\linewidth} | >{\centering\arraybackslash}p{0.15\linewidth} | >{\centering\arraybackslash}p{0.15\linewidth}  | >{\centering\arraybackslash}p{0.15\linewidth} ||}
Structure parameters &	13s & 7s & 8s \\
\hline\hline
Grating period, $a_G$ & 1.0~$\mu$m & 1.5~$\mu$m & 2.5~$\mu$m \\
\hline
Gated region width, $L_G$ & 0.5~$\mu$m & 0.9~$\mu$m & 1.8~$\mu$m\\
\hline
Ungated region width, $L_O$ & 0.5~$\mu$m & 0.6~$\mu$m & 0.7~$\mu$m\\
\hline
Filling factor, $f$ &0.50 & 0.60 & 0.72\\
\hline
Active area &$1.7 \times 1.7$~mm$^2$&$1.7 \times 1.7$~mm$^2$& $1.7 \times 1.7$~mm$^2$ \\
\hline
Number of grating cells, $N_{GC}$ & 1650 & 1100 & 660 \\
\end{tabular}
\end{ruledtabular}
\end{table}

\subsection{Electrical transport characterization}

Before the grating deposition, the bare AlGaN/GaN heterostructures were characterized by capacitance-voltage measurements at 10 kHz using a mercury probe in order to estimate the ungated 2DEG concentration. The extracted value of $n_{O}$ was of $8.7 \pm 0.9 \times 10^{12}$ cm$^{-2}$ at room temperature.

\begin{figure}[t!]
\includegraphics[width=0.45\textwidth]{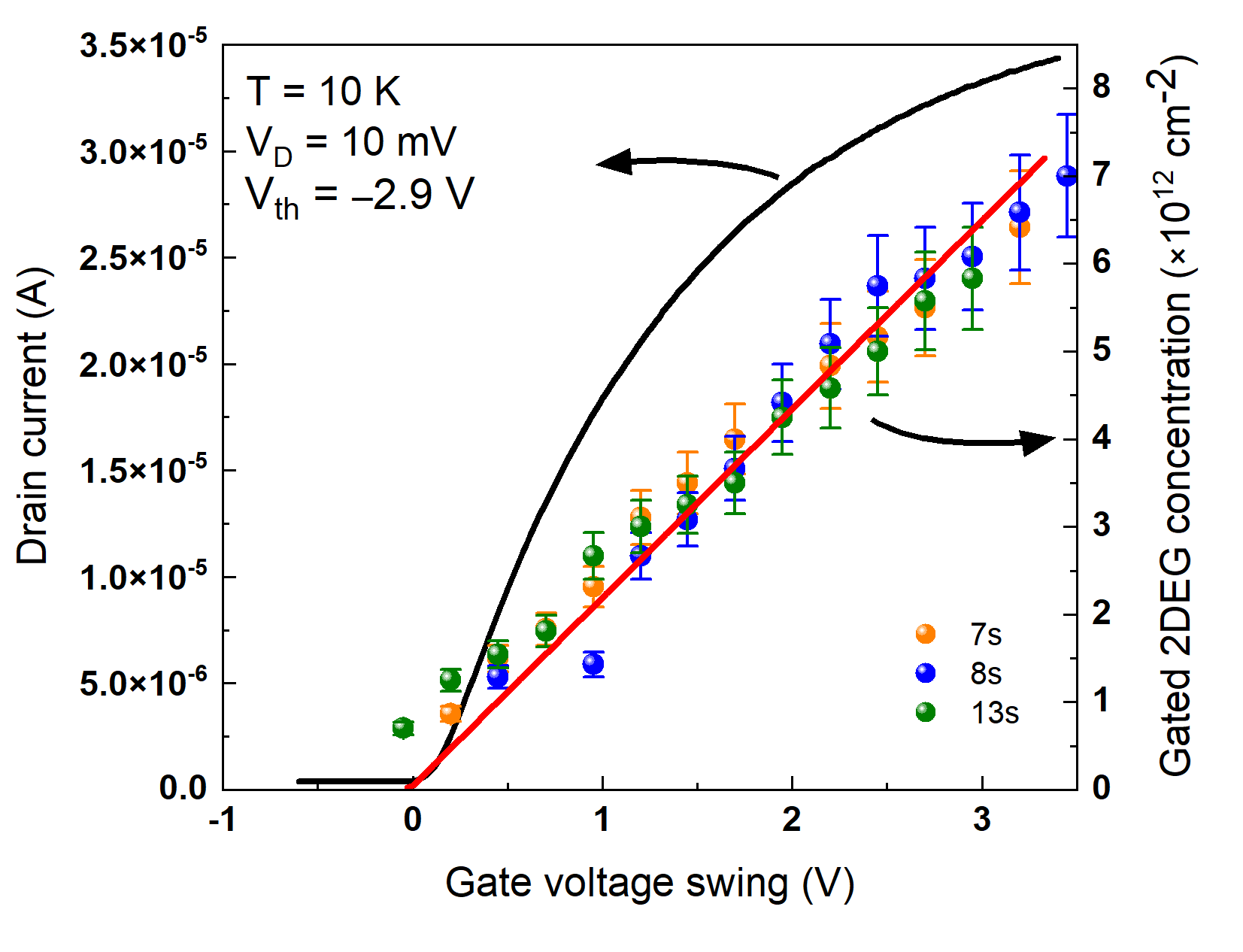}
\caption{\label{Fig5} Transfer current-voltage characteristic of the investigated PC structure measured at 10~K and drain-to-source voltage $V_D = 10$~mV. The red line shows $n_G$ concentration calculated using Eq.~(\ref{capacitor}). Carrier densities $n_G(V_0)$ obtained from the fitting procedure of THz measurements are shown as data points.}
\end{figure}

The transfer current-voltage characteristic of one of the grating-gate samples at $T = 10$~K is shown in Fig.~\ref{Fig5}. It is characterized by the relatively high on \slash off ratio and small subthreshold current, which is an indication of the small gate leakage current. The threshold voltage was determined from the linear extrapolation of the current-voltage characteristic to zero current. For all investigated structures this value was in the range of $V_{th} = -2.9 \pm 0.3$~V. We found $V_{th}$ is only weakly dependent on temperature. Therefore, the carrier density was virtually temperature independent.

\begin{figure}[b!]
\includegraphics[width=0.45\textwidth]{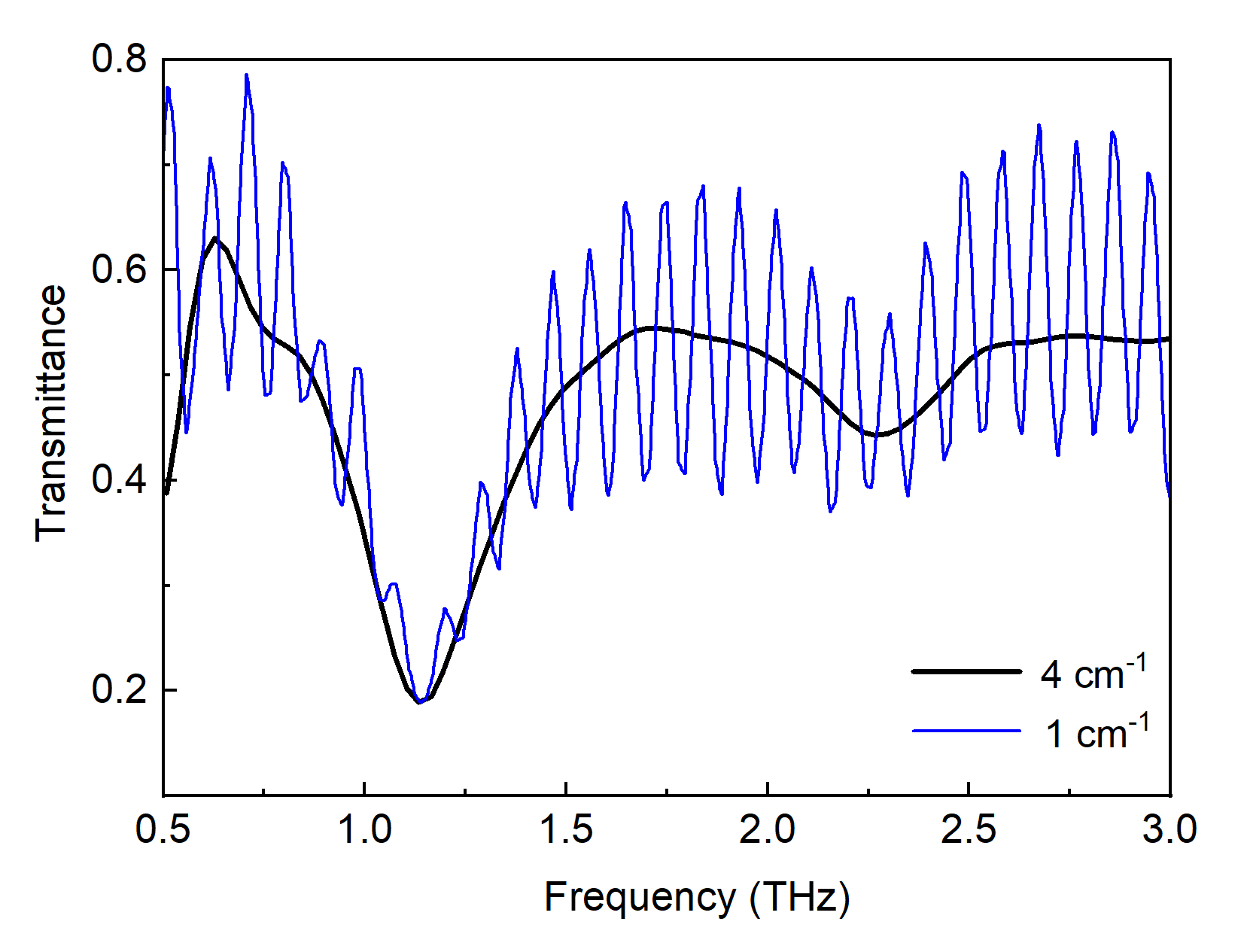}
\caption{\label{Fig6} Experimental transmittance of 7s PC sample at $V_G = 0$~V ($V_0 = 2.9$~V) measured at different spectral resolutions.}
\end{figure}

The 2DEG concentration under the gate, $n_{G}$, for the above threshold regime can be found from the transfer current-voltage characteristics using Eq.~(\ref{capacitor}). This yielded the value of $6.2 \pm 0.6 \times 10^{12}$~cm$^{-2}$ at $V_0 = 2.9$~V. The $n_{G}(V_0)$-dependence is shown in Fig.~\ref{Fig5} by the red line. The DC electron mobility of 2DEG in the fabricated samples was estimated as $\mu = [g_{m0} L_G N_{GC}] / [C W (V_D - I R_{acc})]$, where $g_{m0}$ is the intrinsic transconductance calculated taking into account the access resistance $R_{acc}$ consisting of contact resistances and the resistance of ungated regions, $I$ is current, $V_D$ is drain voltage, $C$ is the gate capacitance per unit area, $N_{GC}$ is a number of grating cells (see Table II), and $W = 1.7$~mm is the total width of the sample. This method provides  $\mu = 7500$~cm$^{2}$/V$\cdot$s at 10~K.

\subsection{THz measurements}

THz transmission spectra measurements of the bare AlGaN/GaN heterostructure as well as PC structures were performed by Fourier-transform infrared (FTIR) vacuum spectrometer (Vertex 80v from Bruker, Billerica, Massachusetts, USA) integrated with continuous flow liquid helium cryostat. In order to reduce optical losses, the original cryostat windows were replaced by polymethylpentene (TPX) windows. The FTIR spectrometer was equipped with a mercury lamp source, a solid-state silicon beam splitter, a cryogenically cooled silicon bolometer, and a 3~THz low pass filter. The experimental spectral range was limited to 3~THz because at higher frequencies the strong phonon absorption of the SiC substrate~\cite{Tarekegne2019} interferes with the 2DEG plasmonic resonance spectra.

FTIR measurements were taken in fast scanning mode at a mirror movement frequency of 5~kHz, with an interferogram average of 100~scans. 
Spectroscopy measurements were taken with a 1.5~mm aperture positioned close to the grating structure, allowing \textit{em} radiation to transmit only through the grating-gate active region. 
A polypropylene film-based linear polarizer was also used in front of the sample. 
In Fig.~\ref{Fig6} we show an example of the transmittance spectra measured at two different resolutions 1~cm$^{-1}$ and 4~cm$^{-1}$. 
The spectra registered at the higher resolution show Fabry-Pérot (FP) fringes, which are caused by multiple internal reflections from the optically-thick 500~$\mu$m SiC substrate.  
In most of the experiments, we used the lower resolution, for which the FB interference is smeared out leaving well-visible plasmonic resonances.
For more details see Supplemental Material~\cite{suppl}: Peculiarities of THz spectra fitting.
All measurements were conducted at 10 K.

\section{Electrical control of plasmonic crystal spectra}\label{Sec4}

\begin{figure}[t!!!]
\includegraphics[width=0.45\textwidth]{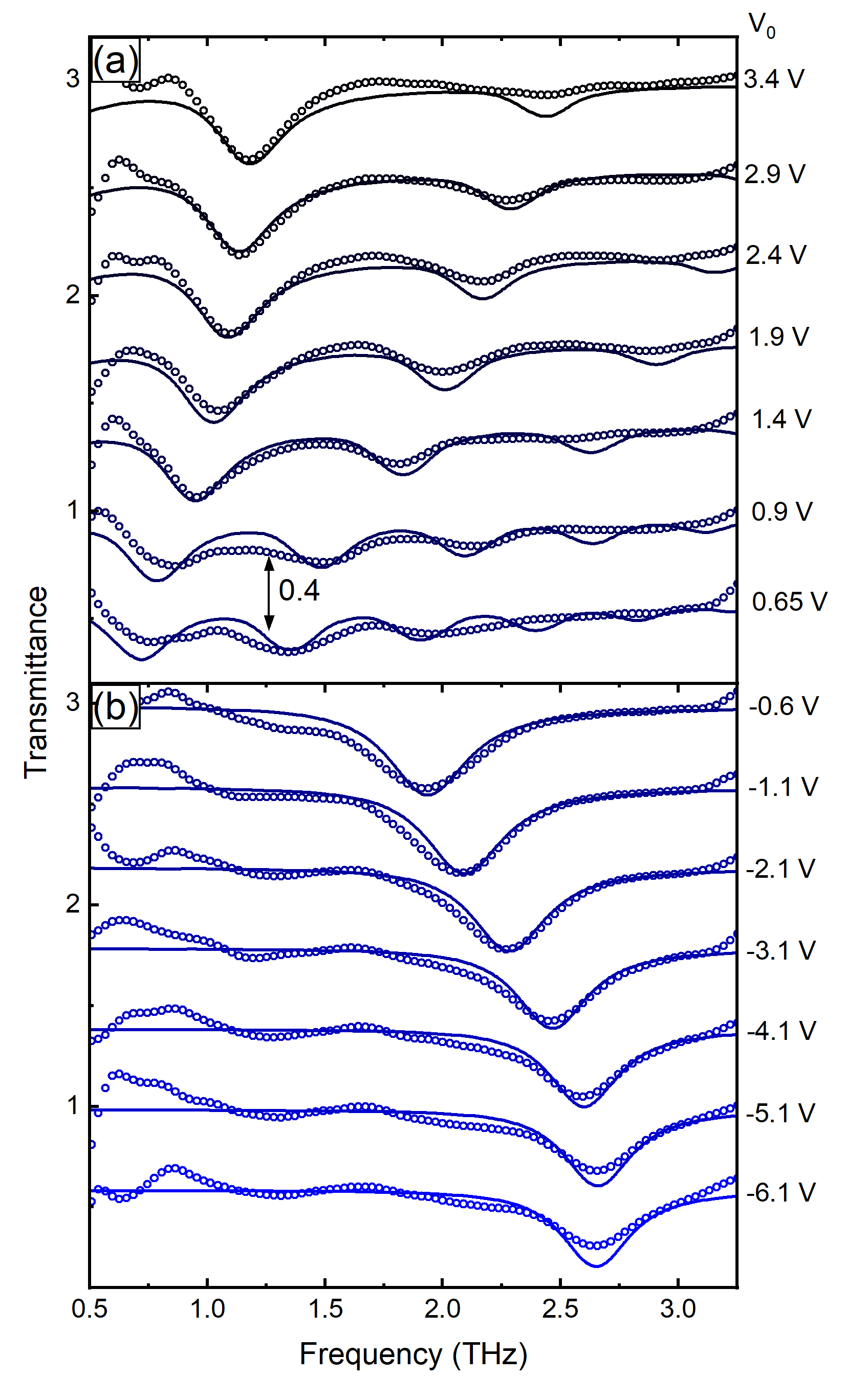}
\caption{\label{Fig7} Transmission spectra of the 7s PC sample at different gate-to-channel voltages: $V_0 > 0$~(a), and $V_0 < 0$~(b). Dotted and solid curves are the results of measurements and calculations, respectively.}
\end{figure}

As was pointed out in Section~\ref{Sec2}, the resonant structure of the transmission spectra is strongly dependent on the modulation degree of 2DEG. The latter can be controlled by the application of the gate voltage.   
Fig.~\ref{Fig7}~(a) shows experimental and calculated transmission spectra of the grating-gated 7s PC sample in the range of gate-to-channel voltages corresponding to modulation degree $\rho < 1$. 
This particular regime is referred to the \textit{ delocalized} phase of the PC.

\begin{figure}[t!!!]
\includegraphics[width=0.45\textwidth]{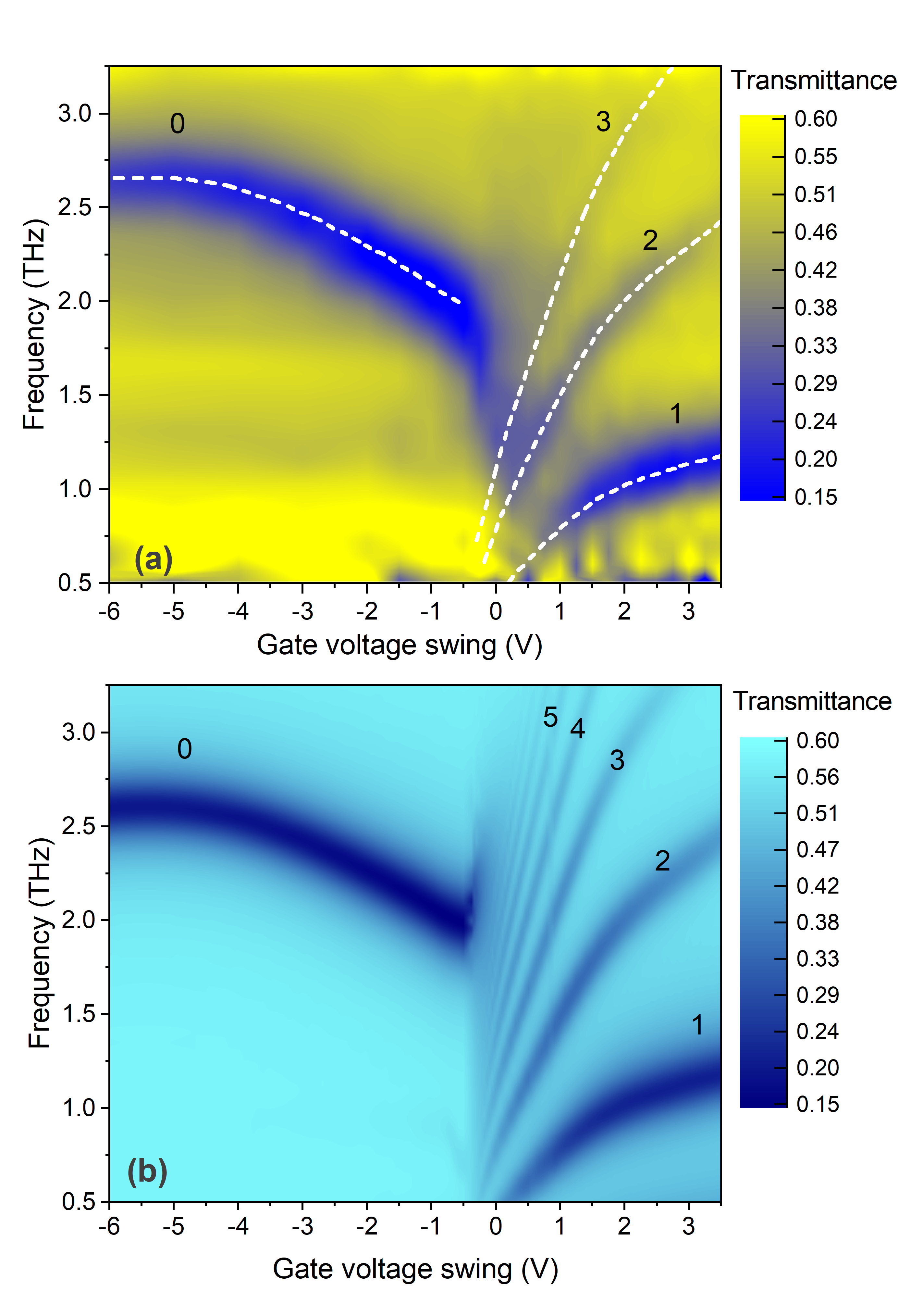}
\caption{Experimental (a) contour plot of the transmittance, $T_{\nu}$, and theoretical (b) contour plot of the transmittance $<T_{\nu}>_{FP}$ of the 7s sample. Dashed curves in panel (a) show the positions of the transmission minima extracted from the calculations.}
\label{Fig8}
\end{figure}

As can be seen in Fig.~\ref{Fig7}~(a), at $V_0 = 3.4$~V two resonances at 1.18~THz and 2.83~THz exist within the experimental frequency band. With the decrease of the gate voltage swing, one can see the gradual red-shift of these two main plasmon resonances ($k = 1, 2$) and the emergence of higher-order resonances. As the gate voltage is approaching the threshold value the resonance structure tends to vanish.
Generally, with a decrease of the $V_0$, the number of resonances increases, but they become less intensive. At $V_0 = 0.65$~V, the calculations predict the emergence of 5 resonances, but only the first three resonances can be seen in the experiment. 

In the sub-threshold regime, $V_0 < 0$, the \textit{localized} phase of the PC is realized with only one resonance in the considered frequency range (see Figs.~\ref{Fig7}~(b) and~\ref{Fig8}~(a)).
With a decrease of $V_0$, we observe a significant blue shift of discussed resonance from frequency 1.93 THz (at $V_0 = -0.6$~V) to 2.64~THz (at $V_0 = -6.1$~V). 
This is an unexpected behaviour because at this range there are no carriers in the gated range and THz radiation is absorbed mainly in the ungated parts. As seen in Fig~\ref{Fig3}~(a) the simple theoretical model predicts in this case that plasma resonance frequency should be gate voltage independent.
We interpret this behaviour as due to the additional shrinking of the ungated region width, $L^{2D}_{O}$, as a result of the gate voltage applied between the gate and the ungated part of the channel. 
Such modification of the channel profile has been reported for the AlGaN/GaN FET with side gates~\cite{cywinski2018, sai2019algan}.

In the fitting procedure for sub-threshold voltages, we varied $\Delta x$ 
in order to account for  the  decrease of width $L^{2D}_{O} = a_G - L_G-2\Delta x$ at constant value of $n_O$ and $n_G = 0$.
Note,  hereafter, all theoretical curves relating to transmission spectra are presented in terms of the quantity $<T_{\nu}>_{FP}$ which is the transmission coefficient calculated for an optically thick substrate with following averaging over a period of FP fringes. 

Figure~\ref{Fig8} shows the experimental contour plots of the transmission spectrum, $T_{\nu}$, in the plane {$V_0 - \nu$} and the calculated one, $<T_{\nu}>_{FP}$, for 7s PC sample. Dashed lines in panel (a) show the calculated frequency of the transmission minima.

The plots in Fig.~\ref{Fig8} allow us to follow the dependencies of the plasmon resonances vs frequency and gate voltage. As seen in Fig.~\ref{Fig8}~(b), calculations predict the existence of 4-5 resonances in \textit{delocalized} PC phase. 
Two of them can be clearly identified in the corresponding experimental contour plot shown in Fig.~\ref{Fig8}~(a). 
In calculations, at $V_0 \geq 0.9$~V, the lowest frequency resonance is the most intensive and corresponds to the fundamental excitation of the 2D plasmon resonance with $k = 1$. 
The resonances enumerated as 2, and 3 are realized in the higher frequencies but they are less intense. 

As seen in Fig.~\ref{Fig8}~(a) and (b), in the narrow region of the near-threshold voltages, the \textit{ localized} plasmon phase of 2DEG strip-grating associated with ungated regions starts to form. At below threshold voltages, from $V_0 = -0.5$~V to $V_0 = - 6$~V, an increase in the frequency of this resonance with saturation at higher gate voltages is observed. 
The general features of the experimental  contour plot (Fig.~\ref{Fig8}~(a)) are well traced by the calculation (Fig.~\ref{Fig8}~(b)).
Particularly, the frequencies of the main resonances and their gate voltage dependencies are well reproduced by the simulations.

\begin{figure}[t]
\includegraphics[width=0.45\textwidth]{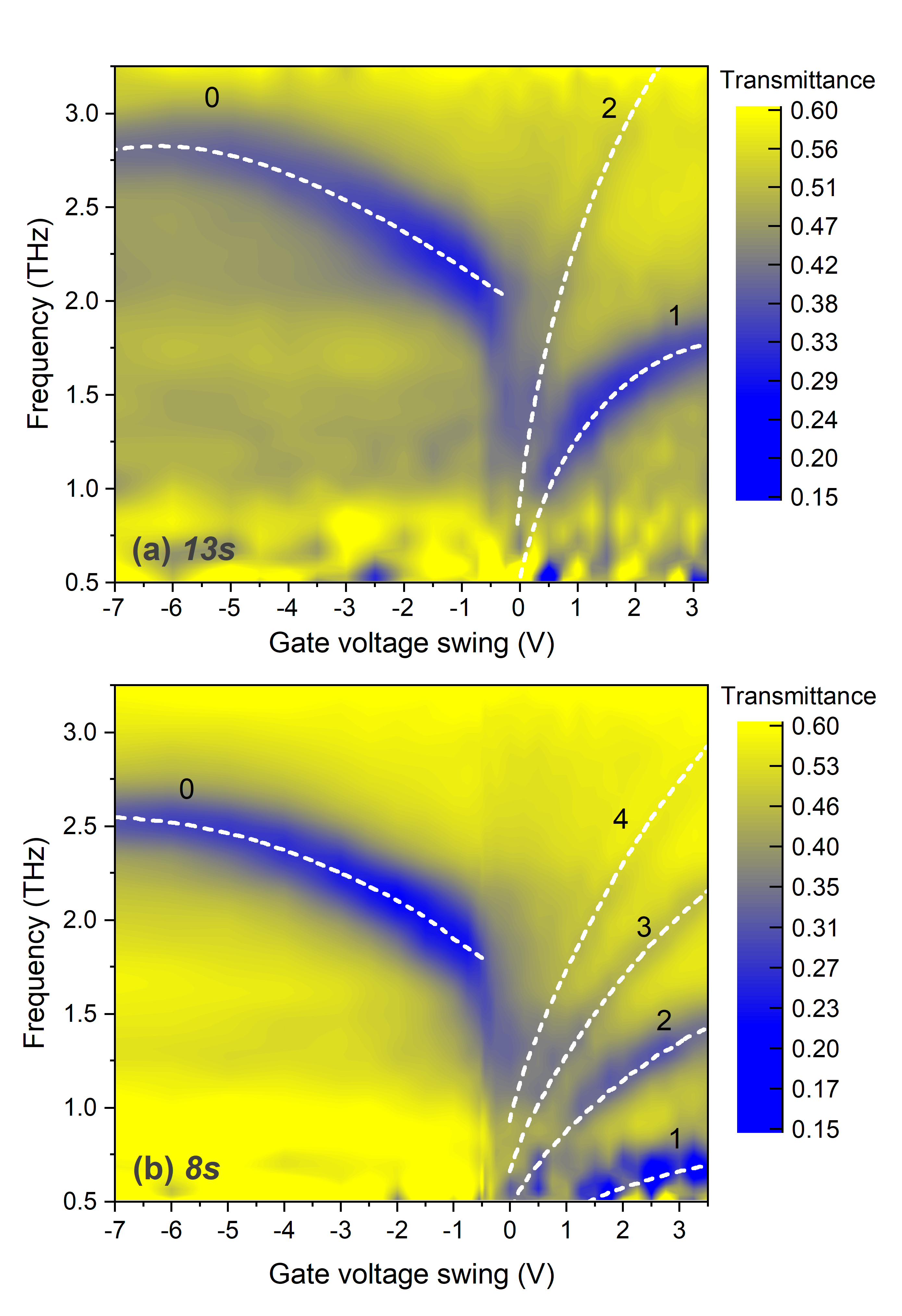}
\caption{Experimental contour plots of the transmittance, $T_{\nu}$ of the 13s (a) and 8s (b) samples. Dashed curves show the positions of the transmission minima extracted from the calculations.}
\label{Fig9}
\end{figure}

For completeness, we present the same experimental contour plots for 13s and 8s PC samples in Fig.~\ref{Fig9}~(a),~(b).
For these samples, the general resonant structure of transmission spectra, their evolution versus gate voltage, and the transition between plasmon resonance regimes are also well reproduced by simulations (dashed lines show the calculated resonance frequencies).

\begin{figure}[b!!!]
\includegraphics[width=0.37\textwidth]{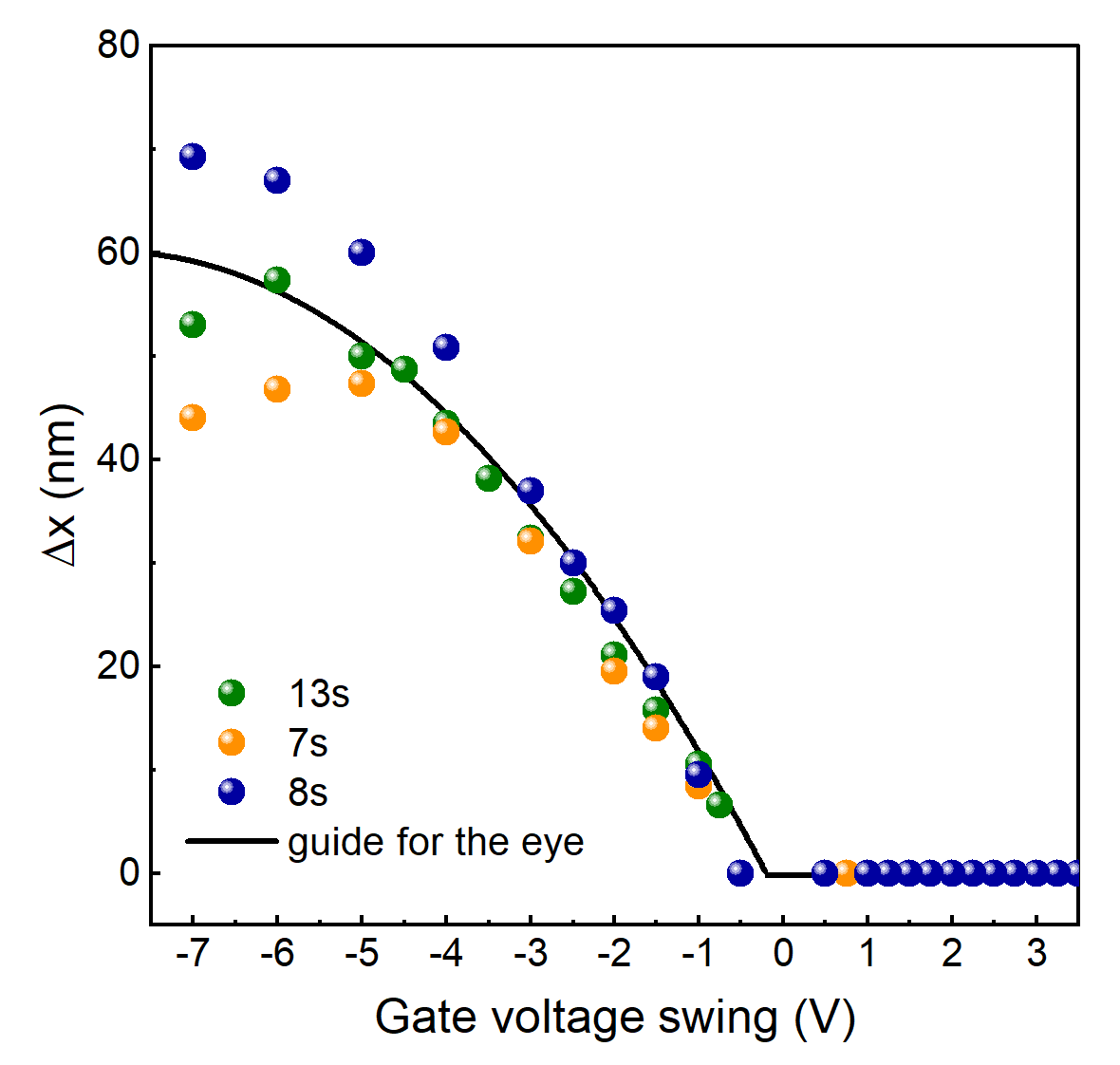}
\caption{
The gate voltage dependencies of the additional depletion length $\Delta x(V_0)$ (dots), the solid black line is shown as a guide for the eye.}
\label{Fig10}
\end{figure}

However, the sharpness of both plasmon phases: \textit{ delocalized} and \textit{ localized} become less pronounced for 13s PC sample which has the smallest filling factor, $f = 0.5$. It is a general property of the grating-gated PC structures for which the wider-slit metallic gratings provide the weaker coupling between \textit{em} wave and plasmon oscillations of 2DEG. 
Moreover, due to the symmetry of the metallic grating of this sample ($L_0 = L_G$), the plasmon modes in uniform 2DEG channel with $k = 2, 4, ...$ are very weakly excited (this peculiarity was pointed out in Ref.~\cite{Michailov1998, KorRCWA2021}). 
Thus, in the experiment at $V_0 > 0$, we can clearly resolve only the fundamental mode of 2D plasmon resonance.
In comparison with 7s PC sample, due to the smaller grating period and the smaller grating filling factor, the resonant frequencies of 13s PC sample are shifted to the higher frequencies. For example, at $V_0 = 2$~V the resonant frequency is equal to 1.57 THz (1.04~THz for 7s sample at the same voltage swing). Because of the smaller width of the ungated region (0.5~µm), the \textit{ localized} plasmon resonance is also at a slightly higher frequency.

For 8s PC sample ($a_G = 2.5$ $\mu$m, $f = 0.72$), the resonant picture of the transmission spectra (see Fig.~\ref{Fig9}~(b)) is very close to the 7s PC sample. At least, three first resonances are well resolved in the experiments and well-reproduced by theoretical calculations. 
However, the larger grating period and larger filling factor of the 8s sample provide resonances at lower, in comparison with 7s and 13s samples, frequencies.
At $V_0 < 1.5$~V, the resonant frequency of the fundamental resonance is out of our experimental spectral range. Because of the larger width of the ungated region (0.7 $\mu$m), the \textit{ localized} plasmon resonance is also realized in the lower frequency range. Therefore, the resonance for gate voltage swing $V_0 = -5$~V is observed at 2.45~THz.
For all three samples, the gate voltage dependencies of the \textit{localized} and \textit{delocalized} plasmon frequencies are well reproduced by the simulations.

In the fitting procedure, we used carrier density in the gated region $n_G$ as the fitting parameter.
The obtained values are shown in Fig.~\ref{Fig5} together with the results of the electrical transport.
One can see that the values used in simulations are very close to those extracted from DC measurements.

Figure~\ref{Fig10} illustrates the relationship between the gate voltage used in simulations and the resulting change in $\Delta x$. It is evident that the dependencies of $\Delta x$ versus $V_0$ exhibit similar behaviour across all three samples. 
Although the width of the ungated region undergoes a certain level of alteration, the observed increase does not exceed 10\% in any of the samples. 
Thus, the change in the width of the ungated region is relatively small.

\section{Discussion and analysis}\label{Sec5}

\subsection{ \textit{Delocalized} phase of plasmonic crystal ($V_0 > 0$)}

\begin{figure}[t!!!]
\includegraphics[width=0.45\textwidth]{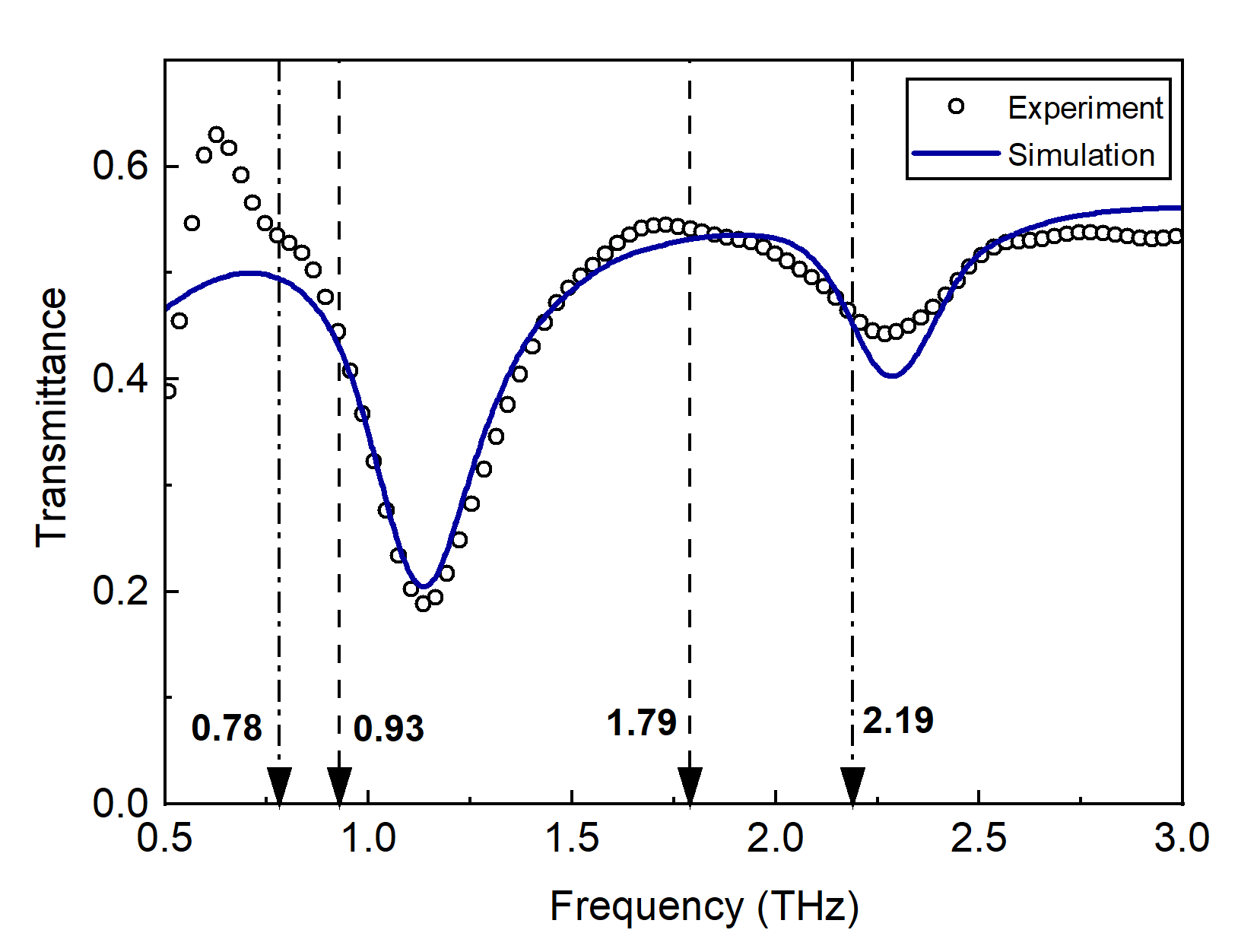}
\caption{\label{Fig11} Experimental, $T_{\nu}$, and simulated, $<T_{\nu}>_{FP}$, transmittance of 7s PC sample at $V_G = 0$~V ($V_0 = 2.9$~V). Vertical arrows show the positions of the first and second-order gated plasmon resonances calculated under the single cavity assumption (dash-dotted lines) and assuming the wave vector defined by the grating period (dashed lines).}
\end{figure}

To better understand the importance of plasmonic crystal approximation let us consider in more details the experimental transmission spectra for the structure 7s at $V_0 = 2.9$~V reproduced once again in Fig.~\ref{Fig11}.
The solid line in this figure shows the result of rigorous electrodynamic simulation. As seen, the simulation very well reproduces the experimental data, including the position of the first and second-order resonances. The best fit was obtained at electron concentrations in the ungated region $n_O = 8.0 \times 10^{12}$ cm$^{-2}$, and in the gated region $n_G = 5.5 \times 10^{12}$ cm$^{-2}$ (modulation degree, $\rho = 0.19$). 
For DC mobility of 2DEG, we choose the value obtained from electrical characterization (see Section~\ref{Sec3}~(B)). 
The values of the parameters obtained from fitting (for more details see Supplemental Material~\cite{suppl}) of the optical spectrum are in good agreement with the values obtained from electrical characterization (see Section~\ref{Sec3}~(B)).

\begin{figure}[b!]
\includegraphics[width=0.45\textwidth]{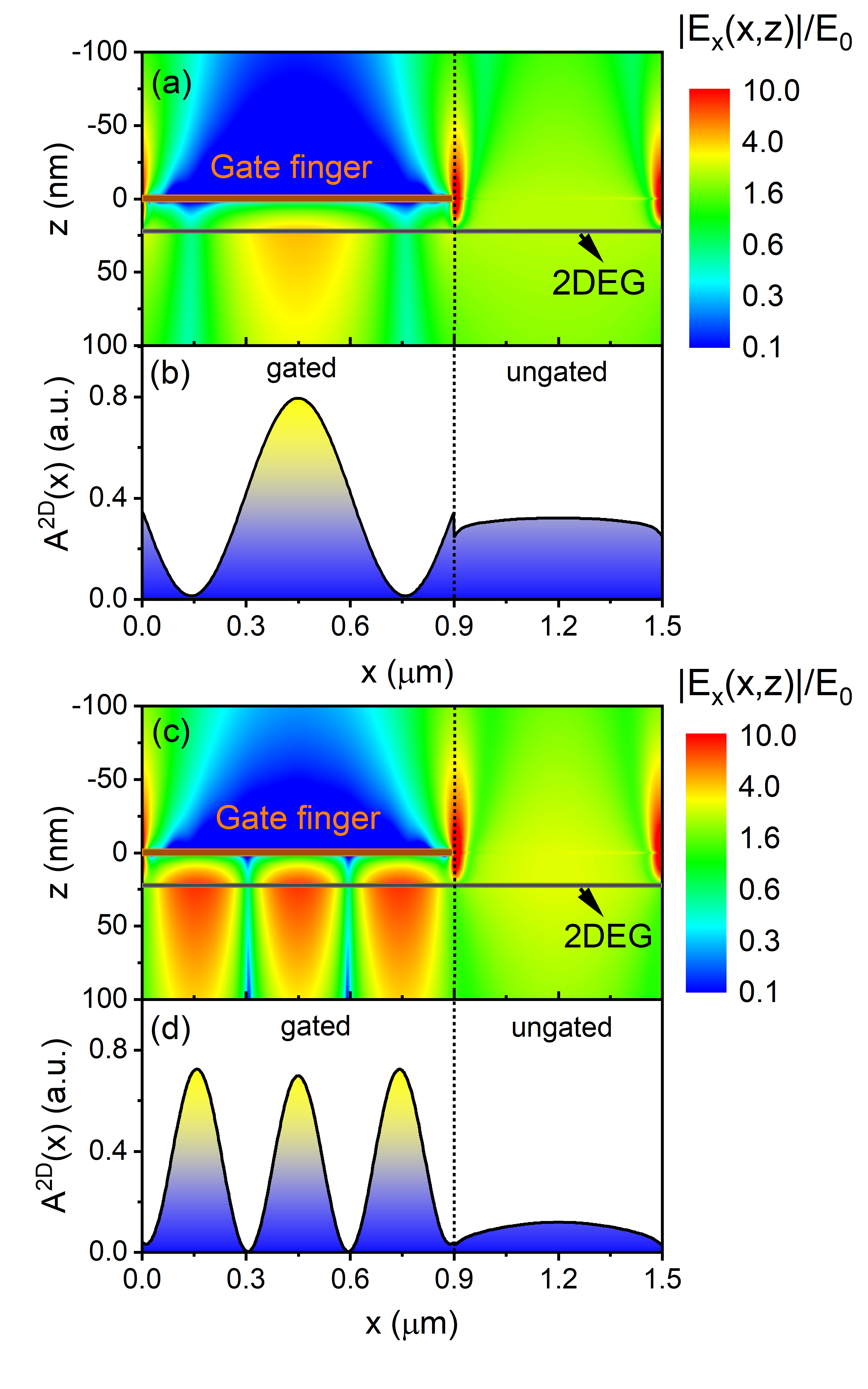}
\caption{
Spatial distributions of the amplitude of $E_x$-component of $em$ field (a, c) and local absorptivity (b, d) in one period of PC at $V_G = 0$~V ($V_0 = 2.9$~V). Results in (a, b) panels are calculated for fundamental plasmon resonance at 1.2~THz. (c, d) panels correspond to the second-order resonance at 2.4~THz (see black stars in the contour plot in Fig.~\ref{Fig3}~(a)).
}
\label{Fig12}
\end{figure}

\begin{figure}[b!]
\includegraphics[width=0.45\textwidth]{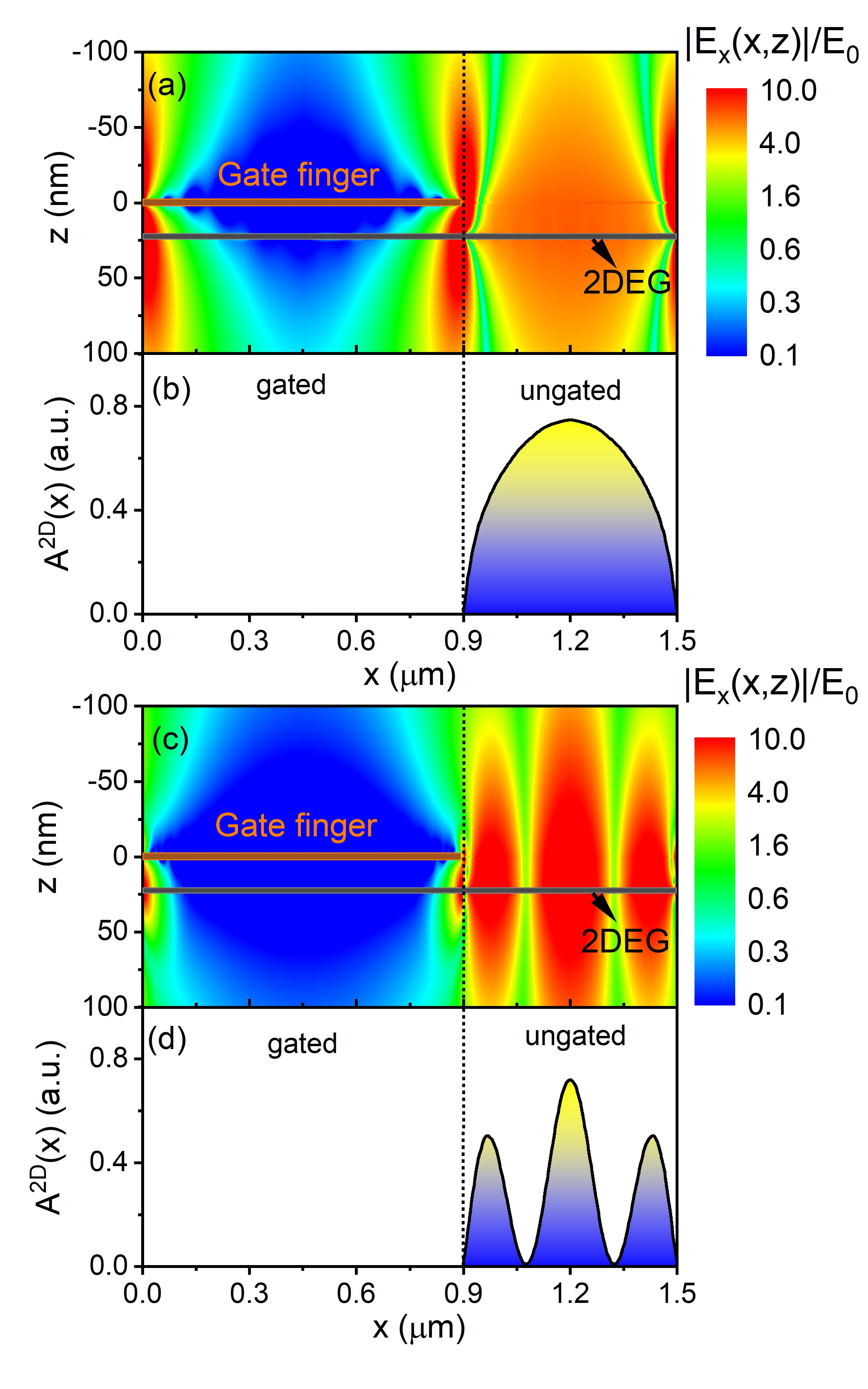}
\caption{
Spatial distributions of the amplitude of $E_x$-component of $em$ field (a, c) and local absorptivity (b, d) in one period of PC at $V_G = -3$~V ($V_0 = -0.1$~V). Results in (a, b) panels are calculated for the fundamental plasmon resonance at 2.3~THz. (c, d) panels correspond to the second-order resonance at 6.1~THz (see red stars in the contour plot in Fig.~\ref{Fig3}~(a)).
}
\label{Fig14}
\end{figure}

In this section, we also aim to compare our model with other simpler approaches. One such method involves estimating the frequencies of the plasmon absorption lines through analytical means. Specifically, we can assume a single cavity description of plasmons and estimate the resonant frequencies. Under this approximation, the grating-gate structure's unit cell breaks down into ungated and gated sections.
These sections can work as cavities with symmetric boundaries for ungated and gated plasmons, separately. Their resonant frequencies can be estimated using Eq.~(\ref{2Dplasmons}) (with corresponding effective permittivities) taking:
\begin{equation}
\label{ungated_k}
q_k=(2k-1)\pi/L_O,\,\, \text{and}\,\, n=n_{O} 		
\end{equation}
for the ungated plasmons localized under the grating openings, and
\begin{equation}
\label{gated_k}
q_k=(2k-1)\pi/L_G, \,\, \text{and}\,\, n=n_{G} 				
\end{equation}
for the gated plasmons localized under the grating strips.

The frequencies of the first and second harmonics for the gated plasmons are shown in Fig.~\ref{Fig11} by dash-dotted lines. 
As can be seen, this estimate is very far from the experimental situation.  
In some publications, the plasmon frequencies were calculated assuming the wave vector defined by the grating period $q_{k}=2\pi k/a_{G}$. 
The first and second-order frequencies for the gated plasmons obtained in this approach are shown by the vertical dashed lines in Fig.~\ref{Fig11}. 
One can see that the simple analytical approaches give results far from the experimental ones and in any case can not be used to describe the grating-gate structure, which should be considered as a PC with strong interaction of the gated and ungated parts.

To get a better understanding of the physical situation we performed the numerical calculations of the electric field profile and absorptivity.
Figure~\ref{Fig12} shows the calculated near-field patterns for the first (Fig.~\ref{Fig12}~(a)) and second (Fig.~\ref{Fig12}~(c)) harmonics for the 7s structure at $V_G = 0$~V ($V_0 = 2.9$~V).
 Particularly, we pay attention to the spatial distributions of the $x$-component of the electric field, $E_x(x,z)$ of the \textit{em} wave in the vicinity of the metallic grating and 2DEG. This component can be used for the calculation of local absorptivity of 2DEG, $A^{2D}$~\cite{KorRCWA2021}.
\begin{equation}
\label{localab}
A^{2D}(x)=\frac{4\pi\text{Re}[\sigma^{2D}(x)]}{c}\times\frac{|E_{x}(x,z_{1})|^2}{\int\limits_{0}^{a_{G}}|E_{0}|^2dx},
\end{equation}
where  $\text{Re}[\sigma^{2D}(x)]$ is the real part of the conductivity profile of 2DEG and $z_{1}$ is the $z$-coordinate of 2DEG.

As seen in Fig.~\ref{Fig12}~(a) and (c), the \textit{em} field in the near-field zone of the considered plasmonic structure has a strongly non-uniform distribution. The field concentration effect is clearly seen near the ridges of the metallic fingers. Also, it is observed the essential increase of the amplitude, $|E_x|$, in the gated region of 2DEG (4 times larger than the amplitude of the incident wave, $E_0$). The latter is associated with the excitation of the fundamental (panel (a)) and second-order (panel (c)) plasmon resonances. The spatial distribution of both amplitude $|E_x(x)(x,z)|$ and local absorptivity $A^{2D}(x)$ (see panels (b) and (d) in Fig.~\ref{Fig12}) acquire oscillating-like behaviour in the gated region with almost flat distribution in the ungated region of 2DEG. 
The discontinuity of quantity $A^{2D} (x)$ at $x=L_G$ is the result of the assumed model of the abrupt step-like concentration profile of 2DEG.

Since these calculations show that the \textit{em} energy absorbs both, in the gated and ungated regions, we can indeed call this regime \textit{delocalized} plasmonic crystal.

\subsection{ \textit{Localized} phase of plasmonic crystal ($V_0 < 0$)}

\begin{figure}[b!!!]
\includegraphics[width=0.45\textwidth]{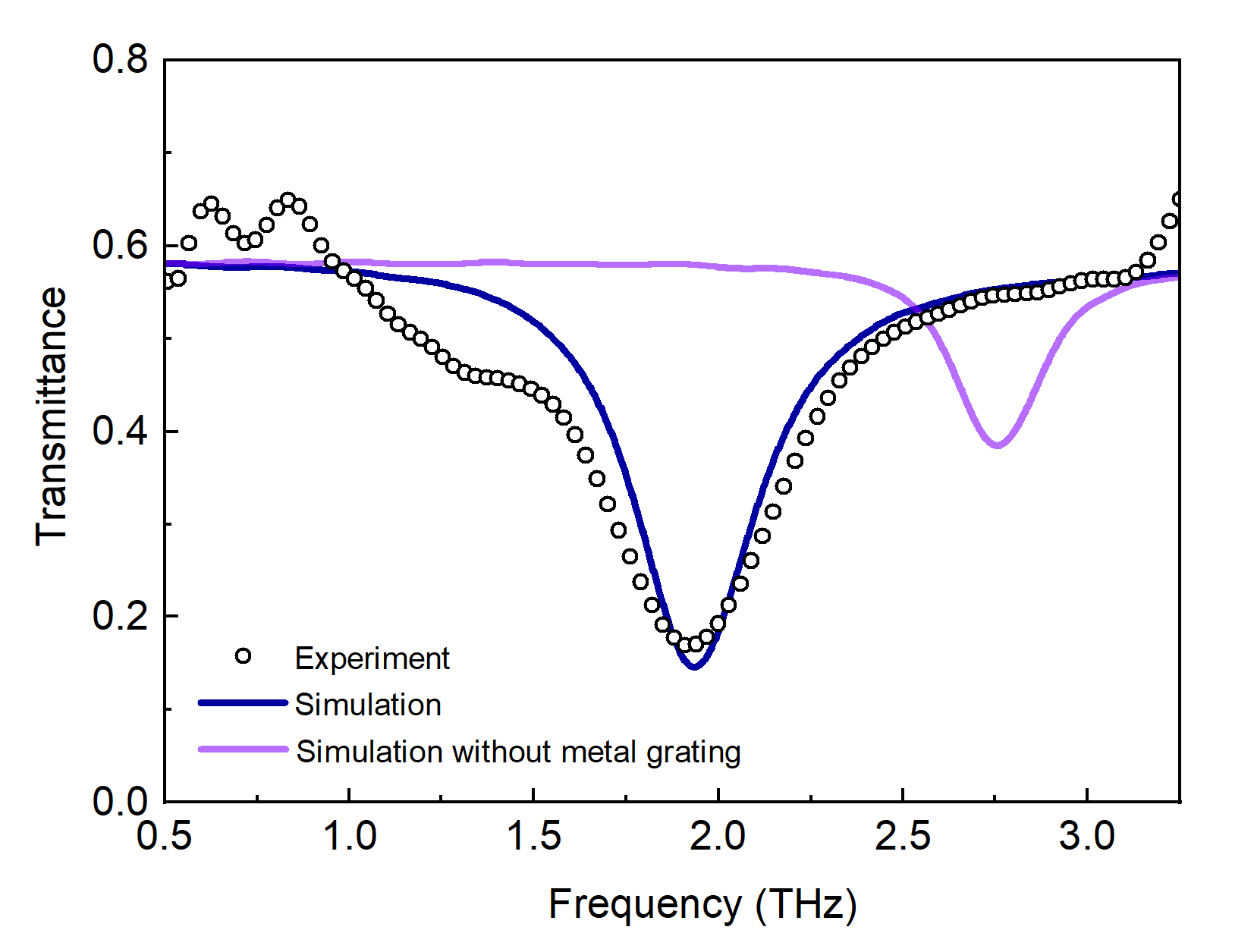}
\caption{\label{Fig13} Experimental, $T_{\nu}$, and simulated, $<T_{\nu}>_{FP}$, transmittance of 7s PC sample at $V_G = -3.5$~V ($V_0 = -0.6$~V).}
\end{figure}

Figure~\ref{Fig14} shows the results of near-field calculations for \textit{localized} phase of PC. In contrast to the \textit{delocalized} plasmon crystal situation (previous section), the energy concentration effect becomes larger. At this, the energy of the \textit{em} wave is mainly concentrated in the region of the metallic grating opening. 
Particularly, the hot zone (high electric field) occurs near the edges of metallic fingers and strips of the formed 2DEG grating. Note that the \textit{em} wave does not penetrate practically under the metallic gate where the cold zone is formed.  
The spatial distributions of the $x$-component of \textit{em} wave in the plane of 2DEG and the local absorptivity (see panels (b) and (d) in Fig.~\ref{Fig14}) show that in this regime ungated part of 2DEG works as a resonant cavity. Particularly, for fundamental resonance (2.3 THz), the plasmon oscillation has a half-wavelength localization on the width of the grating opening.  For the second-order resonance (6.12 THz), the three half-wavelengths localization is clearly seen. As seen in Fig.~\ref{Fig14}, the \textit{ em} energy absorbs only within gate openings. Thus, we call this regime as \textit{ localized} phase of PC.

The near-field calculations emphasize the main physical differences between \textit{ delocalized} and \textit{ localized} phases. For the \textit{ delocalized} phase, the effective absorption of \textit{ em} energy occurs in both, gated and ungated regions of 2DEG. For the \textit{ localized} phase, the plasmon is locally excited in the ungated regions of 2DEG and only these areas work as absorbers of THz waves.

The experimental transmission spectra for the structure 7s at $V_0 = -0.6$~V are reproduced once again in Fig.~\ref{Fig13}. 
As seen, the experiment and simulations are in good agreement. 
We can again attempt to estimate the frequency of the plasmon absorption line analytically using the simple approach. 
The frequency of the ungated plasmon fundamental resonance can be estimated using Eq.~\ref{ungated_k} giving $\nu_{r, 1} = 3.9$~THz, which is far beyond the experimental and simulated values.
In order to understand the role of the metal grating we performed the simulation of the structure neglecting the metallic gates, i.e, the structure consisting only of the ungated parts separated by region with zero electron concentration. 
The result is shown in Fig.~\ref{Fig13} by the violet line. 
As seen, this approach doesn’t give the correct description of the experimental situation. 

Considering different simplified approaches we conclude that even in the absence of carriers under the gated regions, the plasma oscillations of the system cannot be simplified as a sum of independent plasmonic oscillators in the ungated sections. Therefore, the plasmon resonances can only be accurately described using the full PC approximation.

\section{Conclusions}\label{Sec6}

We have experimentally and theoretically investigated resonant properties of large area ($1.7 \times 1.7$ mm$^{2}$) grating-gate AlGaN/GaN plasmonic crystal structures in the THz frequency range. 
Advanced nanometer grating-gate technology allowed us to investigate THz plasma excitations in a wide range of gate voltages, thereby exploring various regimes of carrier density modulation.

The results of the THz measurements were analyzed by rigorous electrodynamic simulations including the contour plots of the \textit{em} field in the near-field zone of the PC structure. The computational algorithm was based on the numerical solution of Maxwell's equations within the method of the integral equations. 
The developed method exhibited a high efficiency with respect to convergence and computational time.

We found that the experimental and calculated resonant structure of the transmission spectra including its modification vs gate voltage corresponds to the formation of two different phases of  PC structure depending on the 2DEG carrier density modulation profile.

Particularly, at above-threshold voltages, $V_0>0$, and small modulation degree $\rho\leq 0.2$, the well-developed \textit{ delocalized} phase of PC is realized. This phase is associated with several resonances in transmission spectra corresponding to different orders of 2D plasmon excitations under the grating. The spatial distribution of the electric field associated with these plasmon excitations (as well as the local absorptivity of THz radiation) extends over the whole period of the structure, i.e. 2D plasma is simultaneously excited in both ungated and gated regions of 2DEG. 

With further gate voltage decrease and approaching the threshold voltage, the transition between \textit{ delocalized} and \textit{ localized} phases of PC starts to form.
At this, the multi-resonant structure in transmission spectra disappears,
higher order harmonics with frequencies resonant with \textit{ localized} plasmons are amplified and finally, they transform into a new strong \textit{ localized} phase.
In contrast to the \textit{ delocalized} phase, the \textit{ localized} phase has spatial localization of the electric field (as well as the local absorptivity of THz radiation) in the ungated part of 2DEG. 
However, the metal grating still plays an important role and the whole structure behaves like a PC. Even if almost the whole \textit{ em} energy is absorbed in the ungated parts the experimental results cannot be explained using only single cavity approaches. 
This is due to the fact that one always has to consider the presence of the metallic grating that partially screens the plasmon oscillations in the 2DEG strips and enhances the coupling of this plasmon resonance with incident radiation.

The general picture of the transformation of PC states from \textit{ delocalized} to \textit{ localized} phases was traced in THz experiments for all three samples with different geometry of the grating coupler. Moreover, we observed that a decrease of $V_G$ below the threshold ($\rho\sim 1$) leads to the essential blue shift of the resonance in the \textit{ localized} phase of the PC structure. 
This phenomenon was explained by the reduction of the ungated part's effective width by negative gate voltage below the threshold (side/edge gate effect).

Our research presents the first experimental evidence of electrically tunable transitions between different phases of THz PCs. The remarkable congruence between our experimental data and calculated values supports the theoretical approach and affirms the physical understanding of all observed phenomena. While our experiments focused on grating-gate structures based on AlGaN/GaN heterojunctions, the findings hold general relevance and can be extended to other semiconductor-based PC structures.
Therefore, our experimental and theoretical results are crucial steps towards a deeper understanding of THz plasma physics and the development of all-electrically tunable devices for THz optoelectronics.

\begin{acknowledgments}
This work was supported by the  "International  Research Agendas" program of the Foundation for Polish Science is co-financed by the European Union under the European Regional Development Fund for CENTERA Lab (No.~MAB/2018/9). The partial support was provided by the National Science Centre, Poland allocated on the basis of Grants No.~2019/35/N/ST7/00203 and No.~2020/38/E/ST7/00476. Also, the work was partially supported by Science and Technology Center in Ukraine (STCU) in collaboration with the Institute of Magnetism of NAS of Ukraine and MES of Ukraine (Project No.~9918). SMK is grateful to Technical University of Dortmund, where Comsol calculations were performed during his PostDoc.  The authors would like to express their gratitude to Prof.~V.~A.~Kochelap (ISP NASU) and Prof.~A.~E.~Belyaev (ISP NASU) for their valuable interest in this work.
\end{acknowledgments}

\nocite{*}

\providecommand{\noopsort}[1]{}\providecommand{\singleletter}[1]{#1}%

\clearpage

\end{document}